\documentclass[
    reprint,
    superscriptaddress,
    nofootinbib,
    amsmath,
    amssymb,
    aps,
    prd
    ]{revtex4-2}

\usepackage{fix-cm}

\usepackage{graphicx}
\usepackage{dcolumn}
\usepackage{bm}
\usepackage{newtxtext,newtxmath}
\usepackage[T1]{fontenc}
\usepackage{amsmath}
\usepackage{glossaries}
\glsdisablehyper
\usepackage{xspace}
\usepackage{multirow}
\usepackage{verbatim}
\usepackage{bm}
\usepackage{hyperref}
\usepackage[dvipsnames]{xcolor}
\usepackage[normalem]{ulem}

\DeclareRobustCommand{\VAN}[3]{#2}
\let\VANthebibliography\thebibliography
\def\thebibliography{\DeclareRobustCommand{\VAN}[3]{##3}\VANthebibliography}

\newcommand{\hfid}{h_{\rm fid}}
\newcommand{\knl}{k_{\rm NL}}

\newcommand{\omc}{\omega_{\rm c}}

\newcommand{\omb}{\omega_{\rm b}}

\newcommand{\wo}{w_0}
\newcommand{\wa}{w_a}
\newcommand{\Mnu}{M_{\nu}}

\newcommand{\ns}{n_{\rm s}}

\newcommand{\As}{A_{\rm s}}
\newcommand{\Omk}{\Omega_{K}}
\newcommand{\omk}{\omega_{K}}
\newcommand{\sigtwelve}{\sigma_{12}}

\newcommand{\Plin}{P_{\rm lin}}

\newcommand{\Mpc}{{\rm Mpc}}
\newcommand{\Mpch}{{\rm Mpc}/h}

\newcommand{\kMpch}{h\,{\rm Mpc}^{-1}}

\newcommand{\lcdm}{$\Lambda$CDM\xspace}
\newcommand{\kcdm}{$K$CDM\xspace}
\newcommand{\wcdm}{$w$CDM\xspace}
\newcommand{\wowacdm}{$\wo\wa$CDM\xspace}
\newcommand{\bone}{b_1}
\newcommand{\btwo}{b_2}
\newcommand{\gtwo}{\gamma_2}
\newcommand{\gtwoone}{\gamma_{21}}
\newcommand{\czero}{c_0}
\newcommand{\ctwo}{c_2}
\newcommand{\cfour}{c_4}
\newcommand{\cnlo}{c_{\rm nlo}}
\newcommand{\npzero}{N^P_0}
\newcommand{\nptwozero}{N^P_{20}}
\newcommand{\nptwotwo}{N^P_{22}}

\newcommand{\Euclid}{{\it Euclid}\xspace}
\newcommand{\Planck}{{\it Planck}\xspace}

\newcommand{\Pell}{P_{\ell}}
\newcommand{\kv}{{\bm k}}

\newcommand{\Sigmav}{{\bm \Sigma}}
\newcommand{\qperp}{q_\perp}
\newcommand{\qpara}{q_\parallel}

\newcommand{\legell}{{\cal L}_\ell}

\newcommand{\Pshot}{P_{\rm sn}}

\newcommand{\kmax}{k_{\rm max}}

\newcommand{\Pgg}{P_{\rm gg}}

\newcommand{\sv}{{\bm s}}

\newcommand{\avir}{a_{\rm vir}}

\newcommand{\Pggoneloop}{P_{\rm gg}^{\,\rm 1\mbox{-}loop}}
\newcommand{\Pggctr}{P_{\rm gg}^{\,\rm ctr}}
\newcommand{\Pggnoise}{P_{\rm gg}^{\,\rm stoch}}

\newcommand{\Thetav}{{\bm \Theta}}
\newcommand{\thetav}{{\bm \theta}}
\newcommand{\nv}{{\bm n}}
\newcommand{\like}{{\cal L}}
\newcommand{\post}{{\cal P}}
\newcommand{\prior}{\Pi}
\newcommand{\Ftwo}{{\cal F}}
\newcommand{\vbar}{\,|\,}
\newcommand{\likestar}{{\cal L}_\star}
\newcommand{\nstar}{{\bm n}_\star}
\newcommand{\nbar}{\bar{n}}
\newcommand{\dv}{{\bm d}}
\newcommand{\tv}{{\bm t}}

\newacronym{pve}{PVE}{prior-volume effects}
\newacronym{des}{DES}{Dark Energy Survey}
\newacronym{desi}{DESI}{Dark Energy Spectroscopic Instrument}
\newacronym{map}{MAP}{maximum a posteriori}
\newacronym{cmb}{CMB}{cosmic microwave background}
\newacronym{eft}{EFT}{effective field theory of large-scale structure}
\newacronym{vdg}{VDG$_\infty$}{velocity difference generating function}
\newacronym{bao}{BAO}{baryon acoustic oscillations}
\newacronym{rpt}{RPT}{renormalised perturbation theory}
\newacronym{dst}{DST}{\emph{discrete sine transform}}
\newacronym{ir}{IR}{infrared}
\newacronym{cdm}{CDM}{cold dark matter}
\newacronym{eds}{EdS}{Einstein-de Sitter}
\newacronym{rsd}{RSD}{redshift-space distortions}
\newacronym{apdist}{AP}{Alcock-Paczynski}
\newacronym{gp}{GP}{Gaussian process}
\newacronym{rbf}{RBF}{radial basis function}
\newacronym{scikit}{SL}{\texttt{scikit-learn}}
\newacronym{lhc}{LHC}{Latin hypercube}
\newacronym{lss}{LSS}{large-scale structure}
\newacronym{fog}{FoG}{finger-of-God}
\newacronym{hod}{HOD}{halo occupation distribution}
\newacronym{pt}{PT}{perturbation theory}
\newacronym{fs}{FS}{full shape}

\newcommand{\pve}{\gls{pve}\xspace}
\newcommand{\des}{\gls{des}\xspace}
\newcommand{\desi}{\gls{desi}\xspace}
\newcommand{\fs}{\gls{fs}\xspace}
\newcommand{\map}{\gls{map}\xspace}
\newcommand{\eft}{\gls{eft}\xspace}
\newcommand{\vdg}{\gls{vdg}\xspace}
\newcommand{\bao}{\gls{bao}\xspace}

\newcommand{\rsd}{\gls{rsd}\xspace}
\newcommand{\apdist}{\gls{apdist}\xspace}

\newcommand{\fog}{\gls{fog}\xspace}

\newcommand{\python}{\texttt{Python}\xspace}

\begin{document}

\preprint{APS/123-QED}

\title{Mitigating prior-volume effects in galaxy clustering by adopting $\Mpc$ over $\Mpch$ units}

\author{Andrea Pezzotta}
\email{andrea.pezzotta@inaf.it}
\affiliation{INAF - Osservatorio Astronomico di Brera, via Brera 28, 20121 Milan, Italy}

\author{Alejandro Pérez-Fernández}
\affiliation{Max Planck Institute for Extraterrestrial Physics, Giessenbachstr. 1, 85748 Garching, Germany}

\author{Giosuè Gambardella}
\affiliation{Institute of Space Sciences (ICE, CSIC), Campus UAB, Carrer de Can Magrans, s/n, 08193 Barcelona, Spain}
\affiliation{Institut d’Estudis Espacials de Catalunya (IEEC), Carrer Gran Capitá 2-4, 08034 Barcelona, Spain}

\author{Ariel G. Sánchez}
\affiliation{Max Planck Institute for Extraterrestrial Physics, Giessenbachstr. 1, 85748 Garching, Germany}

\author{Alexander Eggemeier}
\affiliation{Universität Bonn, Argelander-Institut für Astronomie, Auf dem Hügel 71, 53121 Bonn, Germany}
\affiliation{Cluster of Excellence DYNAVERSE}

\author{Martin Crocce}
\affiliation{Institute of Space Sciences (ICE, CSIC), Campus UAB, Carrer de Can Magrans, s/n, 08193 Barcelona, Spain}
\affiliation{Institut d’Estudis Espacials de Catalunya (IEEC), Carrer Gran Capitá 2-4, 08034 Barcelona, Spain}

\author{Benjamin Camacho-Quevedo}
\affiliation{Institute for Fundamental Physics of the Universe, Via Beirut 2, 34151 Trieste, Italy}
\affiliation{Scuola Internazionale di Studi Superiori Avanzati, via Bonomea 265, 34136 Trieste, Italy}
\affiliation{INAF - Osservatorio Astronomico di Trieste, via Tiepolo 11, 34143 Trieste, Italy}

\date{\today}

\begin{abstract}
We investigate prior-volume effects in Bayesian analyses of galaxy clustering based on state-of-the-art models, such as the EFT and the VDG$_\infty$ model. We show that these effects are largely driven by imposing priors on the model parameters defined in $\Mpch$ units, thereby introducing a spurious correlation with the Hubble parameter $h$, and that they can be strongly mitigated by adopting a more physically motivated formulation in $\Mpc$ units. We test this approach using a set of synthetic data vectors designed to reproduce the clustering signal expected in a Stage-IV survey across a range of cosmological models, including \lcdm, \kcdm, \wcdm, and \wowacdm. We find that the resulting constraints are largely free from shifts induced during the marginalisation process, even in the challenging \wowacdm runs, reducing discrepancies with the fiducial values from several standard deviations to well within $1\sigma$. The performance of this approach is comparable to that of alternative mitigation strategies, such as the adoption of invariant priors or the reparametrisation of nuisance parameters in terms of growth and geometrical amplitudes. For the latter, we find that using $\sigtwelve$ in place of the commonly adopted $\sigma_8$ to describe the amplitude of density fluctuations improves the performance of the reparametrisation technique, even at high redshifts, where constraints on the dark-energy equation of state are weaker. We further show that these conclusions hold with the inclusion of additional BAO constraints, and for data sets with reduced statistical power. These results add to the benefits of working with $\Mpc$ units, avoiding an explicit dependence on $h$, with direct relevance for the analyses of Stage-IV galaxy surveys.
\end{abstract}

\maketitle

\section{Introduction}

The analysis of the large-scale distribution of galaxies from current and upcoming redshift surveys---such as the \des \citep{AbbAguAla2022}, the \desi \citep{DESI2016, DESI2024VII}, and \Euclid \citep{EUCLID2011, MelAbdAce2024}---provides a powerful probe of the cosmological model, enabling precise measurements of both the expansion history and the growth of structures. Cosmological parameters can be inferred either from localised features such as the \bao, or by exploiting the \fs of density-field correlators, most notably, at leading order, the galaxy power spectrum $P(\kv)$ in Fourier space and the two-point correlation function $\xi(\sv)$ in configuration space. In this context, modern analyses rely on increasingly sophisticated theoretical frameworks, such as the \eft \citep{CarHerSen2012, SenZal2014, PerSenJen2016} and the \vdg model \citep{SanScoCro2017, EggLeeSco2024}, which provide a consistent description of galaxy clustering down to mildly non-linear scales while keeping theoretical systematics under control.

Despite their high precision, these approaches can be affected by an important source of systematic uncertainty: \pve. These arise during the process of marginalisation, when projecting the full multi-dimensional posterior onto cosmological parameters, and depend on both the adopted parametrisation and the prior volume assigned to nuisance parameters. When the posterior significantly deviates from a multivariate Gaussian and nuisance priors are broad, such effects can lead to shifts in marginalised constraints, particularly along weakly constrained directions that dominate the posterior volume \citep{HadWolAzz2023}. This issue is especially relevant for galaxy clustering analyses based on the \eft and \vdg frameworks \citep{CarMorPou2023, DESI_FS}, where high-dimensional parameter spaces are needed to describe galaxy bias, \rsd, and small-scale physics, but are only partially constrained by the data. As a result, \pve represent a significant source of systematic uncertainty for ongoing and upcoming Stage-IV galaxy surveys.

Several strategies have been proposed to mitigate these effects. One approach is the use of data-driven priors, such as Jeffreys priors \citep{Jeffreys1946}, which define invariant measures in parameter space based on the local geometry of the likelihood (see \cite{BonParMcG2026} for a recent application to the DESI DR1 sample). More generally, reparametrisation techniques can align sampling coordinates with physically meaningful directions, reducing artificial volume effects induced by non-linear mappings \citep{ParBonChe2025, TseCarMor2026, BonParMcG2026}. Physically motivated priors based on halo occupation statistics offer a complementary avenue, constraining poorly informed directions using simulation or observationally-based constraints \citep{ZhaBonAmi2025}. Similarly, \cite{ReeZhaZhe2025} proposed to correct for \pve directly at the level of the posterior expectation values, by defining them with respect to a non-flat volume measure rather than altering the prior or the parametrisation itself. Alternatively, the issue of prior choice can be avoided altogether by adopting a frequentist approach: profile likelihood methods reduce sensitivity to prior volume by maximising over nuisance parameters rather than marginalising over them; while long established in particle physics and CMB analyses, these methods have only recently started to be applied to large-scale structure data \citep{GomVal2022, HolHerSim2023, ColPouShe2025, HerFerHei2025, MorZhaBon2025}.

Apart from the profile likelihood, these approaches share a common idea: priors must be chosen carefully to control limitations of standard Bayesian inference. While this may appear to conflict with the Bayesian interpretation of priors, it largely reflects the need for a more suitable parametrisation: indeed, even broad flat priors cannot be considered uninformative, as, when applied to data with limited constraining power, they can induce strong projection effects.

In this context, a naturally optimal parametrisation is offered by the \textit{evolution mapping} approach \citep{Sanchez2020, SanRuiJar2022}. By adopting a parametrisation entirely free from any explicit dependence on the expansion rate $h$, this framework makes it possible to cleanly separate the cosmological parameters into two distinct categories: \textit{shape} parameters, which govern the scale dependence of the initial linear power spectrum, and \textit{evolution} parameters, which control the growth history and amplitude of cosmic structures. The basic idea at the heart of evolution mapping is that two-point correlators of density fluctuations become nearly identical across cosmologies sharing the same shape parameters, 
provided that they are evaluated in $\Mpc$ units and compared at epochs when they share the same clustering amplitude. While this mapping is exact in linear theory, \cite{SanRuiJar2022} showed that it remains a good approximation also in the non-linear regime, with small residual deviations caused by the differences in their structure formation histories. 
On a parallel line of work, this framework has been extended to velocity-field correlators \cite{EspSanBel2024} and halo mass functions \cite{FioRuiSan2025}, and used to construct emulators of both the non-linear matter power spectrum measured from $N$-body simulations \cite{SanRuiRod2026} and the \eft and \vdg galaxy power spectrum models \cite{EggCamPez2022, PezEggGam2025}, which exploit these degeneracies to drastically reduce the dimensionality of the training space.

For the analysis of galaxy clustering statistics, a consequence of this framework is that the characteristic scale $\knl$ at which non-linear distortions become significant remains consistent across these models only when defined in $\Mpc$ units. In modern perturbative frameworks, the effective counterterms and stochastic corrections are fundamentally tied to this physical non-linear scale. If these nuisance parameters are instead expressed in the common $\Mpch$ units, they become artificially coupled to the background expansion rate $h$. It is precisely this artificial $h$-dependence that implicitly modifies the prior volume across the parameter space, driving the projection effects.

In this work, we show that a large fraction of \pve in galaxy clustering analyses can be removed simply by imposing priors on parameters expressed in $\Mpc$ units rather than the traditional $\Mpch$ convention, thereby avoiding their explicit dependence on $h$. We test this strategy on a suite of synthetic full-shape analyses spanning \lcdm, \kcdm, \wcdm, and \wowacdm\footnote{We adopt the CPL parametrisation \citep{ChePol2001, Linder2003} of the dark energy equation of state, i.e. $w(a) = \wo + \wa(1-a)$, where $a(t)$ is the cosmic scale factor, set to 1 at present time. In a \wcdm model, the time-evolving component is switched off by setting $\wa=0$, leading to a constant equation of state, $w=\wo.$} cosmologies, assuming an idealised, full-mission Stage-IV survey configuration with spectroscopic bins spanning the redshift range  $0.9<z<1.8$ and an effective sky area of $15\,000\,{\rm deg}^2$, and show that this parametrisation alone is sufficient to substantially reduce prior-induced shifts, regardless of the specific scheme adopted for the remaining nuisance parameters: discrepancies of several standard deviations with respect to the true cosmology, typically found when adopting $\Mpch$-based priors, are reduced to a level well within the $1\sigma$ confidence interval. As a further confirmation, we show that explicitly fixing $h$ to its fiducial value reproduces the same unbiased result obtained in $\Mpc$ units, representing direct evidence that $h$ is the parameter primarily responsible for the emergence of these effects. We find that the performance of this simple approach is comparable to that of more elaborate mitigation strategies, such as reparametrising nuisance parameters in terms of growth and geometrical amplitudes, or adopting invariant Jeffreys priors, while requiring no additional modelling effort. Such methods may still be useful to address residual projection effects that are intrinsic to the geometry of the galaxy clustering likelihood. These conclusions are not specific to this particular setup: we obtain consistent results when combining the full-shape likelihood with post-reconstruction BAO measurements, when adopting a different galaxy clustering model (\eft and \vdg), and when repeating the analysis on data with reduced constraining power, mimicking the setup from an earlier stage of survey completion. This consistency further reinforces the benefits of adopting the $\Mpc$ unit convention as a standard framework for the analysis of galaxy clustering measurements (see e.g. \citep{Sanchez2020, ForFavGom2025} for additional examples related to the inference of amplitude and cosmological parameters).

This paper is organised as follows. In Sect.~\ref{sec:PVE} we provide a brief overview of the origin of \pve (Sect.~\ref{sec:PVE-origin}), and explicitly focus on the appearance of these effects in galaxy clustering likelihoods (Sect.~\ref{sec:parametrisation}). In Sect.~\ref{sec:likelihood} we describe the synthetic data, priors, and likelihood setup adopted in this work. Our main results are presented in Sect.~\ref{sec:results}, while the conclusion is given in Sect.~\ref{sec:conclusions}.

\section{Prior volume effects}
\label{sec:PVE}

In this section, we first review the general origin of \pve in Bayesian inference and illustrate it with a simple toy model (Sect.~\ref{sec:PVE-origin}). We then show how these effects specifically arise in the context of full-shape galaxy clustering analyses, depending on the parametrisation and choice of units adopted (Sect.~\ref{sec:parametrisation}).

\subsection{Formalism}
\label{sec:PVE-origin}

In Bayesian inference, the posterior distribution for a set of model parameters $\thetav = (\Thetav, \nv)$ can be written as
\begin{equation}
    \post({\thetav}\vbar \dv) \propto \like(\dv\vbar\thetav)\,\prior(\thetav)\,,
\end{equation}
where $\Thetav$ denotes the parameters of interest and $\nv$ a set of nuisance parameters, while $\like(\dv\vbar\thetav)$ is the likelihood of drawing data $\dv$ given the full parameter vector $\thetav$, and $\prior(\thetav)$ is the prior distribution encoding external information on the model parameters. The marginal posterior for $\Thetav$ is then obtained by integrating the full posterior over the nuisance parameters,
\begin{equation}
    \post(\Thetav\vbar \dv) \propto \int{\rm d}\nv \, \like(\dv\vbar\Thetav, \nv)\,\prior(\Thetav,\nv)\,.
\end{equation}
To gain intuition on the structure of this expression, we expand the log-likelihood to second order around its maximum $\log(\likestar)$ in the subspace of the nuisance parameters, i.e. we approximate the likelihood as Gaussian in $\nv$ close to its peak, following the Laplace-approximation approach of \cite{KasTieKad1990, TayKit2010}. Denoting by $\nstar(\Thetav)$ the conditional maximum-likelihood solution, with $\Delta \nv \equiv \nv-\nstar$, one obtains
\begin{equation}
    \like(\dv\vbar\Thetav, \nv) \simeq \likestar(\dv\vbar\Thetav)\,
    \exp\left[-\frac{1}{2}\Delta \nv^T \,\Ftwo(\Thetav)\,\Delta \nv\right]\,,
\end{equation}
where
\begin{equation}
    \Ftwo_{ij}(\Thetav) \equiv \left.\frac{\partial^2}{\partial n_i \partial n_j}\left[-\log \like(\dv\vbar\Thetav, \nv)\right]
    \right|_{\nv = \nstar(\Thetav)}
\end{equation}
is the Hessian of the negative log-likelihood with respect to $\nv$. Writing the joint prior as $\prior(\thetav)=\prior(\nv\vbar\Thetav)\,\prior(\Thetav)$ and assuming it to be separable, i.e. 
$\prior(\nv\vbar\Thetav)=\prior(\nv)$, the dependence on $\Thetav$ factorises out of the integral, while the contribution of $\prior(\nv)$ can be absorbed into the definition of $\Ftwo$,\footnote{For the Gaussian priors typically adopted for the nuisance parameters (see Sect.~\ref{sec:likelihood}), with covariance $\Sigmav$, this amounts to the replacement $\Ftwo \rightarrow \Ftwo + \Sigmav^{-1}$. This is the standard procedure followed when marginalising analytically over the nuisance parameters appearing linearly in the model \citep{AmiLewSen2024}.} leading to
\begin{equation}
    \post(\Thetav\vbar d) \propto
    \likestar(\dv\vbar\Thetav) \,
    \prior(\Thetav) \int {\rm d}\nv \,
    \exp\left[-\frac{1}{2}\Delta \nv^T \,\Ftwo(\Thetav)\,\Delta \nv \right],
\end{equation}
and, finally, the evaluation of the Gaussian integral yields
\begin{equation}
    \post(\Thetav\vbar d) \propto
    \likestar(\dv\,|\,\Thetav)\,
    \prior(\Thetav)\,
    \frac{(2\pi)^{\dim(\nv)/2}}{\sqrt{\det \Ftwo(\Thetav)}}\,.
    \label{eq:post_laplace}
\end{equation}
The last factor encodes the effective volume of the nuisance parameter space compatible with the data. Equivalently, defining $\chi^2 \equiv -2\log\post$, Eq.~\eqref{eq:post_laplace} can be recast as
\begin{equation}
    \chi^2_{\rm marg}(\Thetav) = \chi^2_\star(\Thetav) + \log\left[\det\Ftwo(\Thetav)\right] + {\rm const}\,,
    \label{eq:chi2_marg}
\end{equation}
where $\chi^2_\star(\Thetav) \equiv -2\log\left[\likestar(\dv\vbar\Thetav)\,\prior(\Thetav)\right]$ is the profiled $\chi^2$, and we refer to the second term, $\log\left[\det\Ftwo(\Thetav)\right]$, as the \textit{Laplace term}.\footnote{Here we use the term `profiled' even if, strictly speaking, this quantity is more closely related to a \map estimate than to the standard frequentist profile likelihood: the nuisance parameters $\nv$ are profiled out via their conditional maximum-likelihood solution $\nstar(\Thetav)$, but the prior $\prior(\Thetav)$ on the parameters of interest is still retained, unlike a genuine profile likelihood, which excludes any prior contribution.} In particular, whenever $\det \Ftwo(\Thetav)$ exhibits a non-trivial dependence on $\Thetav$---and hence the Laplace term is non-constant---the marginal posterior acquires an additional weighting driven by the varying effective volume of the nuisance subspace. This can induce considerable shifts in the marginal constraints, if the profile likelihood has a shallow slope across a given direction in the nuisance parameter space---see e.g. \citep{HadWolAzz2023} for a demonstration in the context of the 3x2pt analyses. 

This effect can be mitigated by adopting a non-flat prior on the nuisance parameters that compensates for the volume dependence of the Laplace term. The most natural choice is the one provided by the Jeffreys prior \citep{Jeffreys1946},
\begin{equation}
    \prior_{\rm Jeffreys}(\nv\vbar\Thetav) \propto \sqrt{\det\Ftwo(\Thetav)}\,,
    \label{eq:jeffreys_prior}
\end{equation}
which, when substituted in Eq.~\eqref{eq:post_laplace}, exactly cancels the factor $\det[\Ftwo(\Thetav)]^{-1/2}$ arising from the Gaussian integration over $\nv$. In terms of the marginalised $\chi^2$, the Laplace term in Eq.~\eqref{eq:chi2_marg} drops out, leaving only the profiled $\chi^2_\star(\Thetav)$, which is effectively converging to a frequentist approach.

\begin{figure}
    \includegraphics[width=\columnwidth]{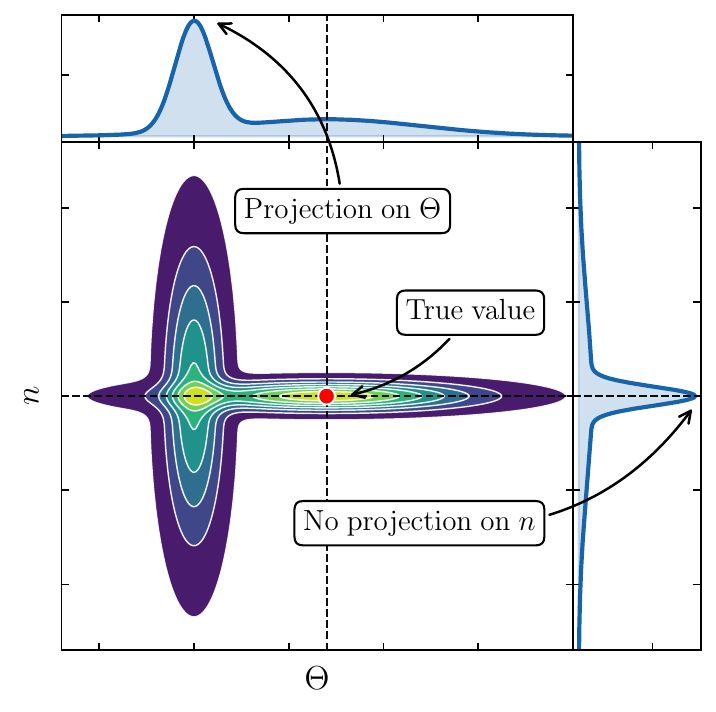}
    \caption{Toy model illustrating the emergence of projection effects when marginalising over nuisance parameters $\nv$ with a strongly varying hypervolume as a function of the remaining parameters $\Thetav$ (see also Fig.~1 of \cite{HadWolAzz2023, GomVal2022} for related illustrations).}
    \label{fig:example_PVE}
\end{figure}

The problem of \pve can nonetheless be visualised intuitively through a simple toy model, illustrated in Fig.~\ref{fig:example_PVE}. Even when the fiducial point---corresponding, for instance, to the maximum-likelihood solution and thus to the maximum of $\likestar(\Thetav)$---is consistent with the full multidimensional posterior, marginalisation over poorly constrained nuisance parameters can lead to a non-uniform contribution from different regions of the nuisance parameter space. As a result, regions with larger effective volume contribute more to the marginal posterior, shifting the inferred constraints on $\Thetav$ away from their fiducial value.

In the context of galaxy clustering analyses, $\Thetav$ typically corresponds to the cosmological parameters of interest, while $\nv$ collects the nuisance parameters describing galaxy bias, \rsd, and small-scale corrections within the \eft and \vdg frameworks introduced in Sect.~\ref{sec:parametrisation}. As we show in Sect.~\ref{sec:results}, the choice of units adopted to express these nuisance parameters, either $\Mpch$ or $\Mpc$, directly affects the effective volume $\det\Ftwo(\Thetav)$ associated with them, making the unit convention itself a primary driver of \pve in this class of analyses.

\subsection{Parametrisation of full-shape models}
\label{sec:parametrisation}

The interpretation of galaxy clustering measurements on mildly non-linear scales relies on increasingly sophisticated theoretical frameworks capable of accurately modelling non-linear gravitational evolution, galaxy bias, and \rsd. Among the most widely used approaches are the \eft and \vdg models, both based on a perturbative description of linear and mildly non-linear modes, supplemented by effective parameters that capture the impact of physics on scales not directly resolved by the perturbative expansion.

In both \eft and \vdg models, the redshift-space galaxy power spectrum can be schematically written as \citep{PerSenJen2016, EggLeeSco2024}
\begin{equation}
\Pgg(k,\mu) = \Pggoneloop(k,\mu) + \Pggctr(k,\mu) + \Pggnoise(k,\mu)\,,
\label{eq:Pgg}
\end{equation}
where $\mu$ is the cosine of the angle between the wave vector and the line-of-sight. Here $\Pggoneloop$ denotes the perturbative contribution up to next-to-leading order (one-loop) arising from the coupling of density and velocity fields, $\Pggctr$ collects effective counterterms accounting for the small-scale \rsd signal and the impact of unresolved small-scale physics beyond the perturbative regime, and $\Pggnoise$ represents stochastic contributions sourced by small-scale density fluctuations.

In each of these contributions, the model depends on a number of nuisance parameters that are typically marginalised over in the inference process. The one-loop term $\Pggoneloop$ depends on a set of bias parameters, including the linear bias $\bone$, the quadratic bias $\btwo$, and the tidal biases $\gtwo$ and $\gtwoone$ (see \citep{EggScoCro2020} for a description of the bias basis adopted in this work).
Matter counterterms $\{c_\ell\}_{\ell=0,2,4}$ contribute to $\Pggctr$ as
\begin{equation}
\Pggctr(k,\mu)\supset -2\,c_\ell\, k^2\Plin(k)\legell(\mu)\,,
\label{eq:ctr_terms}
\end{equation}
where $\Plin$ is the linear-theory power spectrum,\footnote{In the correct expression this should be replaced by the leading-order infrared-resummed power spectrum, to properly account for the damping of the \bao feature induced by large-scale bulk flows \citep{EisSeoSir2007, CroSco2008, BalMirSim2015}. For simplicity, we retain $\Plin$ in this and the following expressions.} and $\legell$ is the Legendre polynomial of order $\ell$. In the \eft model, a further next-to-leading-order contribution must be included explicitly, scaling with the fourth power of $k$ and controlled by the free parameter $\cnlo$ \citep{IvaSimZal2020, EggLeeSco2024}. Conversely, in the \vdg model, the deterministic components are explicitly damped by an analytical function that accounts for small-scale \rsd effects---the \fog effect---and depends on an additional free parameter $\avir$. Similarly, at one-loop order, stochastic corrections receive contributions from the parameters $\{\npzero,\nptwozero,\nptwotwo\}$ of the form
\begin{equation}
\Pggnoise(k,\mu) \supset \nbar^{-1}\,\left[\npzero + N^P_{2\ell} \,k^2 \legell(\mu)\right]\,,
\label{eq:noise_terms}
\end{equation}
where $\nbar$ denotes the mean number density of the galaxy sample under consideration.\footnote{This parametrisation allows for a direct assessment of deviations from Poissonian shot noise, $\Pshot \equiv \nbar^{-1}$. Alternatively, one may sample directly over the overall amplitude of the stochastic contribution (see, e.g., \citep{IvaSimZal2020}).}

Finally, we account for the \apdist effect \citep{AlcPac1979}, which arises from an incorrect choice of fiducial cosmology when converting the observed angles on the sky and redshifts into comoving distances. This effect distorts the transverse and line-of-sight components of the wave vector $\kv$, which can be written as
\begin{equation}
k_{\perp,{\rm fid}} = \qperp\, k_{\perp}\,,\qquad
k_{\parallel,{\rm fid}} = \qpara\, k_{\parallel}\,.
\end{equation}
Here, $\qperp$ and $\qpara$ are the transverse and line-of-sight dilation parameters, defined as
\begin{equation}
\qperp \equiv \frac{D_{\rm M}(z)}{D_{{\rm M, fid}}(z)}\,,\qquad
\qpara \equiv \frac{H_{\rm fid}(z)}{H(z)}\,,
\end{equation}
where $D_{\rm M}(z)$ is the transverse comoving distance and $H(z)$ is the Hubble parameter.\footnote{In the context of the \bao\ measurements discussed in Sect.~\ref{sec:results}, the analogous dilation parameters $\alpha_\perp$ and $\alpha_\parallel$ are instead defined relative to the sound horizon at the drag epoch, $r_{\rm d}$: $\alpha_\perp \equiv \qperp\, r_{\rm d,fid}/r_{\rm d}$ and $\alpha_\parallel \equiv \qpara\, r_{\rm d,fid}/r_{\rm d}$.} Rather than considering the full two-dimensional dependence of the power spectrum on $k$ and $\mu$, it is standard practice to compress the information onto the basis of Legendre polynomials $\legell(\mu)$, where at leading-order the only non vanishing terms come from $\ell = \{0,2,4\}$. The observed power spectrum multipoles are then obtained as
\begin{equation}
    \Pell(k_{\rm fid}) = A_{\rm AP}\,\frac{2\ell+1}{2}\int_{-1}^{+1} \Pgg(k,\mu)\,\legell(\mu_{\rm fid})\,{\rm d}\mu_{\rm fid}\,,
    \label{eq:multipoles}
\end{equation}
where $A_{\rm AP} \equiv (\qperp^2\qpara)^{-1}$ is the volume rescaling factor due to \apdist distortions, and the relations $k(k_{\rm fid},\mu_{\rm fid})$ and $\mu(\mu_{\rm fid})$ can be found in \cite{BalPeaHea1996}.

We note that, contrary to what is customarily assumed in full-shape galaxy clustering analyses, the stochastic contribution $\Pggnoise$ of Eq.~\eqref{eq:noise_terms} does not require the $A_{\rm AP}$ rescaling appearing in Eq.~\eqref{eq:multipoles}.
This factor converts a quantity computed in the model (true) cosmology into the volume element defined by the fiducial cosmology adopted for the data.  However, as the mean number density $\nbar$ entering Eq.~\eqref{eq:noise_terms} is already fixed to its fiducial value during parameter estimation, it inherently accounts for the fiducial volume.
Therefore, no such volume rescaling is needed for the stochastic terms, unlike for the remaining contributions to $\Pgg$. In Appendix~\ref{app:full_contours} we show how this assumption affects the resulting posterior.

With this parametrisation, the leading-order counterterms and scale-dependent shot-noise parameters from Eqs.~\eqref{eq:ctr_terms}--\eqref{eq:noise_terms} carry dimensions of length squared, while $\cnlo$ and $\avir$ carry dimensions of length to the fourth power and length, respectively. Their numerical values therefore depend on the adopted unit convention. A more physically motivated parametrisation of these effects can be obtained by introducing an explicit dependence on characteristic non-linear scales associated with the different contributions. In general, these scales can be defined through the variance of the linear density field smoothed on a scale $R_{\rm NL}$,
\begin{align}
    &\sigma^2(R) = \int \frac{{\rm d}^3k}{(2\pi)^3}\, P_{\rm lin}(k)\, W^2(kR)\,, \\
    &\sigma^2(R_{\rm NL})\equiv 1\,, \qquad \knl \sim R_{\rm NL}^{-1}\,,
\end{align}
where $W(kR)$ is the usual top-hat window function. 
Different physical effects entering the effective description---such as higher-derivative bias, velocity dispersion, or stochastic contributions---are expected to be controlled by distinct characteristic scales, rather than a single universal $\knl$. In this spirit, the counterterm and shot-noise contributions can be schematically reparametrised as
\begin{equation}
p\, k^i \;\rightarrow\; \hat{p} \left(\frac{k}{\knl}\right)^i\,,
\end{equation}
with $p$ denoting a generic dimensional nuisance parameter and $\hat{p}$ its dimensionless counterpart, $i=2$ for the counterterms $c_\ell$ and the noise parameters $N^P_{2\ell}$, and $i=4$ for $\cnlo$.

Even with this reparametrisation, the dependence on the unit convention is not removed but simply transferred from the nuisance parameters to the non-linear scale. This, however, makes the origin of the problem more transparent. As shown by the evolution-mapping framework \citep{SanRuiJar2022, SanRuiRod2026}, when distances are expressed in $\Mpc$, the expansion rate $h$ affects non-linear clustering essentially only through the clustering amplitude, which can be quantified by $\sigtwelve$, the RMS linear density fluctuation in spheres of radius $R=12\,\Mpc$. Cosmologies sharing the same shape parameters and $\sigtwelve$ therefore exhibit nearly identical non-linear clustering, regardless of their value of $h$. The onset of non-linearities---and, with it, the characteristic scales controlling the counterterm and stochastic contributions---therefore occurs at very similar scales in $\Mpc$ across such cosmologies, but at different scales in $\Mpch$. Fixing these scales, or equivalently imposing priors on the corresponding dimensional parameters, in $\Mpch$ units thus artificially couples them to the expansion rate $h$, introducing spurious correlations between the nuisance and cosmological parameters that drive the prior-volume effects during marginalisation.

\begin{table}
\caption{Rescaling factors $(X_{\Mpch} = (\text{factor})\times X_{\Mpc})$ relating quantities expressed in $\Mpch$ and $\Mpc$ units, so as to obtain the same $\chi^2$ when evaluating the likelihood in the two unit systems. The scale, power spectrum multipoles, covariance, and scale cuts are rescaled through the fixed fiducial value $h_{\rm fid}$, while the non-linear scale must be rescaled through the actual value of $h$ at which the likelihood is evaluated. For the number density, the factor outside (inside) parentheses refers to our baseline treatment, where the stochastic terms $\Pggnoise$ are not rescaled by $A_{\rm AP}$ (to the standard approach, where $A_{\rm AP}$ also multiplies the stochastic terms).}
\vspace{6pt}
\centering
\renewcommand{\arraystretch}{1.3}
\begin{tabular}{|c|c|c|}
\hline
Quantity & Notation & Factor \\
\hline
scale & $k$ & $h_{\rm fid}^{-1}$ \\
\hline
observable & $\Pell$ & $h_{\rm fid}^{3}$ \\
\hline
covariance & $C$ & $h_{\rm fid}^{6}$ \\
\hline
non-linear scale & $k_{\rm NL}$ & $h^{-1}$ \\
\hline
number density & $\bar{n}$ & $h_{\rm fid}^{-3}$ $\big(h^{-3}\big)$ \\
\hline
scale cuts & $k_{\rm max}$ & $h_{\rm fid}^{-1}$ \\
\hline
\end{tabular}
\label{tab:unit_conversion}
\end{table}

Despite this, it is always possible to match the $\chi^2$ exactly between the two unit systems for any single likelihood evaluation. This is achieved through a consistent conversion of all relevant quantities—including the data vector, covariance matrix, and model parameters—between $\Mpc$ and $\Mpch$ units, as summarised in Table~\ref{tab:unit_conversion}. Because the underlying physical predictions are identical, the likelihood itself is perfectly invariant under a simple change of units. 

However, Bayesian inference relies on the posterior distribution, which depends not just on the maximum likelihood but on the volume of the parameter space defined by the priors. If the broad priors on the nuisance parameters are set and held fixed in their respective native units, the choice between $\Mpc$ and $\Mpch$ imposes a different prior geometry, which directly alters the Laplace term in Eq.~\eqref{eq:chi2_marg}. Consequently, when marginalising over the expansion rate $h$, the integral averages over distinct effective volumes depending on the unit convention. This means that while a change of variables can perfectly map the $\chi^2$ at a single point, the resulting marginalised posteriors will inevitably differ, driven entirely by the unit-dependent \pve.

\begin{figure*}
    \includegraphics[width=2\columnwidth]{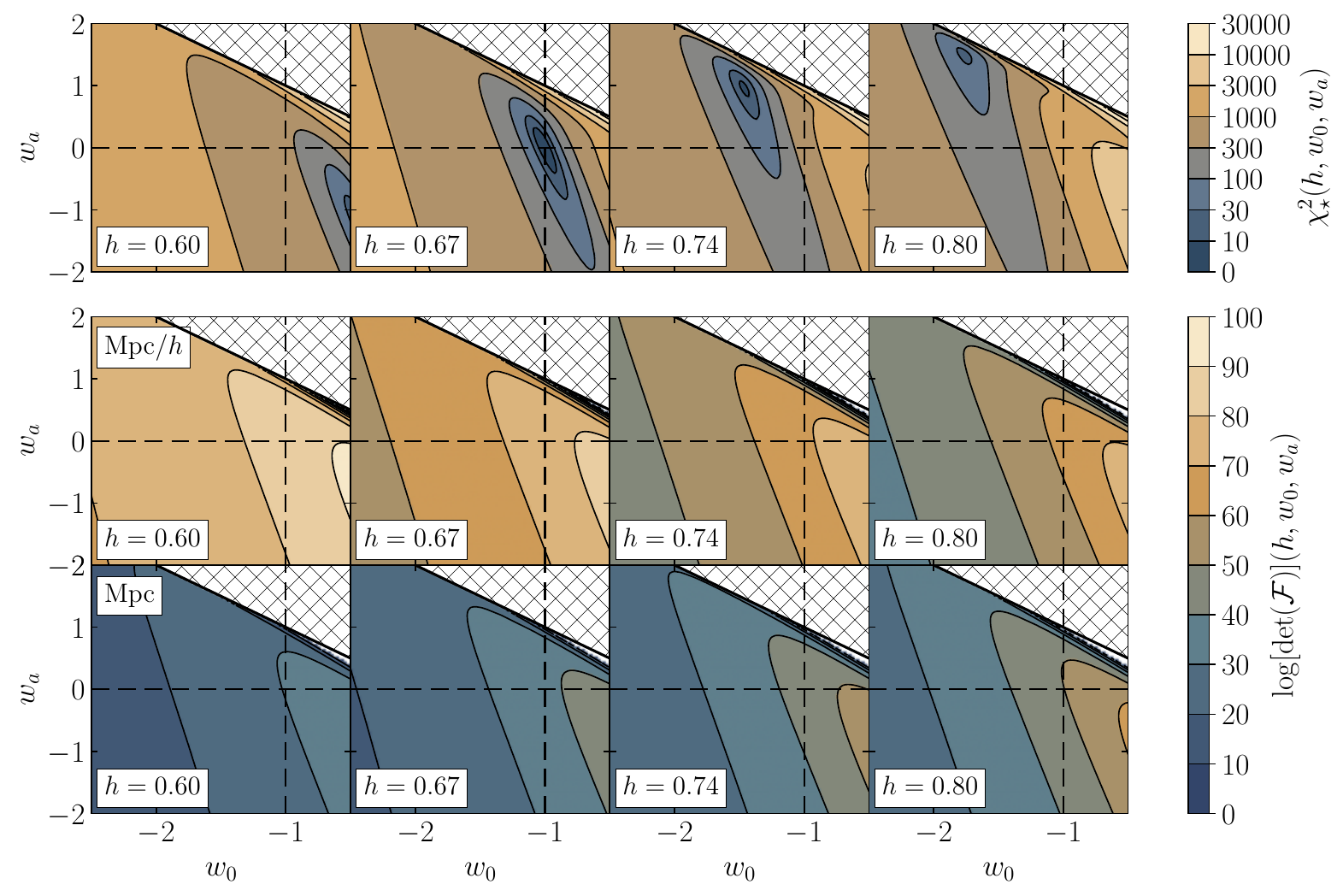}
    \caption{Slices of the profiled $\chi^2$ (top row) and of the Laplace term (middle and bottom rows) in the $(\wo,\wa)$ parameter subspace. The difference between the middle and bottom rows reflects the use of different distance units---$\mathrm{Mpc}/h$ (middle) and $\mathrm{Mpc}$ (bottom)---while different columns correspond to different values of $h$, as indicated in each panel. The fiducial $\Lambda$CDM model is marked by the intersection of the dashed lines. Iso-contours of the profiled $\chi^2$ and of the Laplace term are shown as solid black lines and colored levels, following the scale indicated by the corresponding colorbars. For the sake of comparison, the middle and bottom rows share the same colorbar, allowing a clearer quantification of their relative differences.}
    \label{fig:laplace_term_hw0war}
\end{figure*}

Figure~\ref{fig:laplace_term_hw0war} illustrates this point by showing slices of the profiled $\chi^2_\star$ and of the Laplace term $\log[\det({\cal F})]$ in the $(\wo,\wa)$ parameter subspace, computed using the synthetic data vector and likelihood setup described in Sect.~\ref{sec:likelihood}. For all cases, we show snapshots at different values of $h$ of the corresponding two-dimensional distributions in the $(\wo,\wa)$ parameter space, while keeping all cosmological and non-linear nuisance parameters fixed to their fiducial values, and marginalising the linear nuisance parameters as described in Sect.~\ref{sec:likelihood}. Note that this implies that the resulting $\chi^2_\star$ and Laplace term do not exactly coincide with those obtained when marginalising over the full posterior distribution; nevertheless, even this limiting case already provides useful insight into the origin of \pve.

The behaviour of the profiled $\chi^2$ is as expected. When the correct value of the expansion rate is reached, $h=0.67$, the maximum-likelihood point converges towards the fiducial \lcdm values ($\wo=-1$, $\wa=0$). Conversely, deviations in $h$ can be partially compensated by adjustments in the dark energy equation of state, following a degeneracy direction that anticorrelates $\wo$ and $\wa$. In particular, low (high) values of $h$ require a larger (smaller) amplitude of $w(z)$, as can be seen from the shift of the minimum of the $\chi^2$ across different panels. As the two unit choices lead to identical results, we only show the results from the $\Mpc$ case.

Turning to the Laplace term, we find that even at fixed $h$ its amplitude decreases towards the upper-left region of the panels, indicating a preference for lower values of $\wo$ and higher values of $\wa$. This is clearly reflected in the orientation of the iso-contours. While the variation within individual panels may not be sufficient to significantly impact the marginalised posterior---as shown later in Fig.~\ref{fig:w0wa_fixed-h}---sampling over $h$ induces an artificial drift of the parameter space exploration towards this region, driven by the mismatch between the integrated contributions across different $h$ slices. This effect is markedly more pronounced in $\Mpch$ units, where both the overall amplitude of the Laplace term and its variation with $h$ are larger. In $\Mpc$ units, by contrast, the orientation and amplitude of the Laplace term are nearly insensitive to the specific value of $h$, resulting in much weaker \pve. It is therefore natural to expect that, in a standard analysis, the marginalised posteriors on $\{h, \wo, \wa\}$ will be confined to regions of parameter space displaced from their fiducial values purely as a consequence of this marginalisation effect.\footnote{As shown by \cite{TseCarMor2026}, this bias can be corrected by adopting a suitable reparametrisation of the nuisance parameters, as we also discuss in Sect.~\ref{sec:results}.} As we show in Sect.~\ref{sec:comp_units_priors}, it is precisely this distinction that is ultimately responsible for the different posterior distributions obtained in the two unit systems.

\section{Data vectors, priors, and likelihood setup}
\label{sec:likelihood}

Having established the theoretical origin of \pve and their connection to the choice of units adopted to express full-shape model parameters, we now turn to the data and likelihood setup used to quantify these effects in practice. Since our goal is to detect the presence of \pve, in this work we make use of synthetic data vectors constructed using the same theoretical model employed in the exploration of the posterior distribution. This choice excludes any deficiency of the theoretical model itself---of the kind that can arise when analysing N-body simulations or real observations, where the model only provides an approximate description of the data---ensuring that our results are not contaminated by systematic effects unrelated to \pve.

For this purpose, we construct two sets of synthetic data vectors, corresponding to the \eft and \vdg models, respectively. We assume that these are measured from four non-overlapping redshift bins spanning the range $0.9<z<1.8$, with a sky fraction of $f_{\rm sky}\sim1/3$, mimicking the galaxy population targeted by the Euclid spectroscopic survey. In all cases, our observables are the first three even Legendre multipoles of the anisotropic redshift-space galaxy power spectrum, as defined in Eq.~\eqref{eq:multipoles}, assuming the true cosmology as our fiducial model such that \apdist distortions naturally vanish. For each redshift bin, we adopt fixed scale cuts $\kmax=0.3$, $0.25$, and $0.25\,\kMpch$ for the monopole, quadrupole, and hexadecapole, respectively. When the analysis is performed in $\Mpc$ units, these scale cuts are correspondingly rescaled to $\kmax\times h_{\rm fid}$, consistently with the conversion factor reported in Table~\ref{tab:unit_conversion}. The fiducial cosmology adopted to generate the data vectors is a flat \lcdm model with a minimal neutrino mass, and is reported in Table~\ref{tab:cosmology}. The values of the nuisance parameters describing bias, \rsd, and shot noise are chosen to reproduce a clustering signal representative of a population of H$\alpha$ emitters \citep{CamCroPel2026}; their values are reported in Appendix~\ref{app:fiducial_nuisance}.

\begin{table}
\caption{Fiducial cosmology adopted to generate the synthetic data vectors, corresponding to a flat \lcdm model with a minimal neutrino mass. The table reports the values of the Hubble parameter $h$, the cold dark matter, baryon, and curvature density parameters $\omc$, $\omb$, and $\omk$, the total neutrino mass $\Mnu$, and the amplitude $\As$ and spectral index $\ns$ of primordial scalar perturbations, together with the dark energy equation-of-state parameters $\wo$ and $\wa$.}
  \vspace{6pt}
  \centering
  \renewcommand{\arraystretch}{1.3}
  \begin{tabular}{|c|c|c|c|c|c|c|c|c|}
    \hline
    $h$ & $\omc$ & $\omb$ & $\omk$ & $\Mnu\,[\mathrm{eV}]$ & $\As\,[10^{-9}]$ & $\ns$ & $\wo$ & $\wa$ \\
    \hline
    $0.67$ & $0.121203$ & $0.0219961$ & $0$ & $0.06$ & $2.1$ & $0.96$ & $-1$ & $0$ \\
    \hline
  \end{tabular}
  \label{tab:cosmology}
\end{table}

The covariance matrices for each set of measurements are estimated using a Gaussian approximation \citep{GriSanSal2016}, which is not expected to significantly affect the comparison between different strategies to mitigate \pve. Additionally, we do not include any survey window function, and instead work directly with the unconvolved theoretical predictions obtained from the different \rsd models.

The posterior distribution over the parameter space is obtained using the \texttt{nautilus} nested sampling algorithm \citep{Lange2023}, which draws samples from the prior $\Pi(\thetav)$ and evaluates at each point the log-likelihood which, assuming Gaussianity, reads
\begin{equation}
    \log{\cal L}(\dv\,|\,\thetav) = -\frac{1}{2}\left[\dv-\tv(\thetav)\right]^T C^{-1}\left[\dv-\tv(\thetav)\right]\,,
\end{equation}
up to an additive constant. Here, $\dv$ and $\tv$ denote the data and theory vectors, corresponding to the stacked Legendre multipoles ordered consistently with the covariance matrix $C$. We exploit the possibility of analytically marginalising the nuisance parameters that enter linearly in the different \rsd models, in order to speed up the convergence of the sampler \citep{AmiLewSen2024}. This procedure can be applied to the subset of parameters $\{\gtwoone,\czero,\ctwo,\cfour,\cnlo,\npzero,\nptwozero,\nptwotwo\}$, effectively leaving only the cosmological parameters and $\{\bone,\btwo,\gtwo,\avir\}$ to be explicitly sampled.

The prior ranges adopted for the baseline $\Mpch$ and $\Mpc$ configurations are listed in Table~\ref{tab:priors}; in all cases, the fiducial cosmological and nuisance parameters used to build the data vectors lie well within the explored prior ranges. The priors on the cosmological parameters are flat, with the exception of $\omb$ and $\ns$, for which we adopt Gaussian priors informed by external data: a BBN-based prior on $\omb$ \citep{Schoneberg2024}, and a 10-times $\Planck$ prior on $\ns$ \citep{Planck2018}. This choice mirrors the prior setup adopted in the DESI full-shape galaxy clustering analyses \citep{DESI_FS}. To define a consistent set of priors in $\Mpc$ units, we rescale the bounds (or mean and standard deviation) of each dimensional nuisance parameter by the corresponding power of $h_{\rm fid}$. Since $h$ itself is sampled over, this rescaling does not amount to an exact one-to-one mapping between the $\Mpch$ and $\Mpc$ priors; nevertheless, to first approximation it provides a sensible prescription for relating the two unit systems.

\begin{table}
\caption{Prior ranges adopted for the sampled cosmological and nuisance parameters, in the baseline ($\Mpch$) and $\Mpc$ configurations. The notation for the prior adopts ${\cal U}(a,b)$ and ${\cal N}(\mu,\sigma)$ for uniform and normal distributions, respectively, where $a$ and $b$ are the lower and upper bounds, and $\mu$ and $\sigma$ are the mean and standard deviation. The curvature $\Omk$, and the dark energy equation of state parameters, $w$ and $(\wo,\wa)$, are included as free parameters only in the \kcdm, \wcdm, and \wowacdm extensions, respectively. Similarly $\cnlo$ and $\avir$ are included only in the \eft and \vdg models, respectively. The fiducial value of the Hubble parameter, $h_{\rm fid}$, is used to rescale the priors on the nuisance parameters that are not dimensionless. Priors marked with an asterisk correspond to parameters that are analytically marginalised, while the remaining ones are directly sampled.}
  \vspace{6pt}
  \centering
  \renewcommand{\arraystretch}{1.3}
  \setlength{\tabcolsep}{10pt}
  \begin{tabular}{|c|c|c|}
    \hline
    Parameter & \multicolumn{2}{c|}{Prior} \\
    \hline
    $h$ & \multicolumn{2}{c|}{${\cal U}(0.5, 0.9)$} \\
    \hline
    $\omc$ & \multicolumn{2}{c|}{${\cal U}(0.08, 0.16)$} \\
    \hline
    $\omb$ & \multicolumn{2}{c|}{${\cal N}(0.0219961, 0.00038)$} \\
    \hline
    $\As\,[10^{-9}]$ & \multicolumn{2}{c|}{${\cal U}(1.05, 2.95)$} \\
    \hline
    $\ns$ & \multicolumn{2}{c|}{${\cal N}(0.96, 0.042)$} \\
    \hline
    $\omk$ & \multicolumn{2}{c|}{${\cal U}(-0.15, 0.15)$} \\
    \hline
    $w\,/\,\wo$ & \multicolumn{2}{c|}{${\cal U}(-2.5, 0)$} \\
    \hline
    $\wa$ & \multicolumn{2}{c|}{${\cal U}(-3, 3)$} \\
    \hline
    $\bone$ &  \multicolumn{2}{c|}{${\cal U}(0.25, 4)$} \\
    \hline
    $\btwo$ & \multicolumn{2}{c|}{${\cal U}(-5, 5)$} \\
    \hline
    $\gtwo$ & \multicolumn{2}{c|}{${\cal U}(-4, 4)$} \\
    \hline
    $\gtwoone$ & \multicolumn{2}{c|}{${\cal N}(0, 5)$} \\
    \hline
    $\npzero$ & \multicolumn{2}{c|}{${\cal N}(0, 2)$} \\
    \hline
    & $\Mpch$ & $\Mpc$ \\
    \hline
    $\czero$ & ${\cal N}(0, 100)$ &  ${\cal N}(0, 100 / h_{\rm fid}^2)$\\
    \hline
    $\ctwo$ & ${\cal N}(0, 100)$ &  ${\cal N}(0, 100 / h_{\rm fid}^2)$\\
    \hline
    $\cfour$ & ${\cal N}(0, 100)$ &  ${\cal N}(0, 100 / h_{\rm fid}^2)$\\
    \hline
    $\cnlo$ & ${\cal N}(0, 200)$ & ${\cal N}(0, 200 / h_{\rm fid}^4)$\\
    \hline
    $\avir$ & ${\cal U}(0, 15)$ &  ${\cal U}(0, 15 / h_{\rm fid})$\\
    \hline
    $\nptwozero$ & ${\cal N}(0, 30)$ & ${\cal N}(0, 30 / h_{\rm fid}^2)$\\
    \hline
    $\nptwotwo$ & ${\cal N}(0, 30)$ & ${\cal N}(0, 30 / h_{\rm fid}^2)$\\
    \hline
  \end{tabular}
  \label{tab:priors}
\end{table}

For a subset of these runs, we additionally implement two alternative mitigation strategies, anticipated here for later comparison in Sect.~\ref{sec:results}. We apply Jeffreys priors only to the nuisance parameters entering linearly in the model, $\{\gtwoone,\czero,\ctwo,\cfour,\cnlo,\npzero,\nptwozero,\nptwotwo\}$, for which the Laplace term can be removed analytically from $\chi^2_{\rm marg}$, as in Sect.~\ref{sec:PVE-origin} (see \cite{BonParMcG2026} for a comparison with a full Jeffreys implementation including all nuisance parameters, applied to DESI DR1). For the reparametrisation strategy, we instead adopt a scheme inspired by \cite{TseCarMor2026}, hereafter denoted as TCM or ${\rm TCM_{linear}}$ depending on whether the reparametrisation is applied to the whole set of nuisance parameters or just to the ones appearing linearly in the model. In this setup, each nuisance parameter $p$ is replaced by the rescaled parameter
\begin{equation}
\tilde p = A_{\rm AP}^{\,n_p}\,\sigma^{m_p}\,p\,,
\end{equation}
where $A_{\rm AP}$ is the \apdist volume rescaling factor already defined in Eq.~\eqref{eq:multipoles}, and $\sigma$ denotes a density-fluctuation amplitude: $\sigma_8$, the RMS linear density fluctuation in spheres of radius $R=8\,\Mpch$, when working in $\Mpch$ units, or $\sigtwelve$, when working in $\Mpc$ units. Following \cite{TseCarMor2026}, we adopt $(n_p,m_p)=(1/2,1)$ for $\bone$, $(1/2,2)$ for $\btwo$ and $\gtwo$, $(1,4)$ for $\gtwoone$, and $(1,2)$ for the counterterms $\{\czero,\ctwo,\cfour,\cnlo\}$, while $\avir$ is left unchanged. In our baseline setup, the shot-noise parameters $\{\npzero,\nptwozero,\nptwotwo\}$ are also left unchanged, consistently with the absence of the $A_{\rm AP}$ rescaling of the stochastic contributions discussed in Sect.~\ref{sec:parametrisation}; in the runs of Appendix~\ref{app:full_contours}, where this rescaling is included, we instead adopt $(n_p,m_p)=(1,0)$ for them. For consistency, the priors on the rescaled nuisance parameters are correspondingly rescaled by the appropriate power of $\sigtwelve/\sigma_8$. Conversely, since the distribution of $A_{\rm AP}$ values is consistent with unity, no rescaling of the prior is applied in that case.

\section{Results}
\label{sec:results}

In this section, we compare the marginalised posterior distributions obtained under the different prior and likelihood configurations introduced in the previous sections, with the aim of establishing which setup is most effective at mitigating \pve. We focus in particular on the comparison between priors defined on nuisance parameters expressed in $\Mpch$ units and the corresponding priors defined in $\Mpc$ units, while also assessing how this comparison is affected by the alternative mitigation strategies discussed above.

As discussed in Sect.~\ref{sec:parametrisation} and illustrated in Fig.~\ref{fig:laplace_term_hw0war}, the Laplace term entering the marginalised $\chi^2$ of Eq.~\eqref{eq:chi2_marg} is markedly steeper as a function of $\wo$ and $\wa$ in $\Mpch$ units than in $\Mpc$ units, leading to a systematic drift of the marginalised posterior whenever $h$ is sampled over. We now quantify the impact of this effect on the full Bayesian inference of the cosmological parameters.

\begin{figure*}
    \includegraphics[width=2\columnwidth]{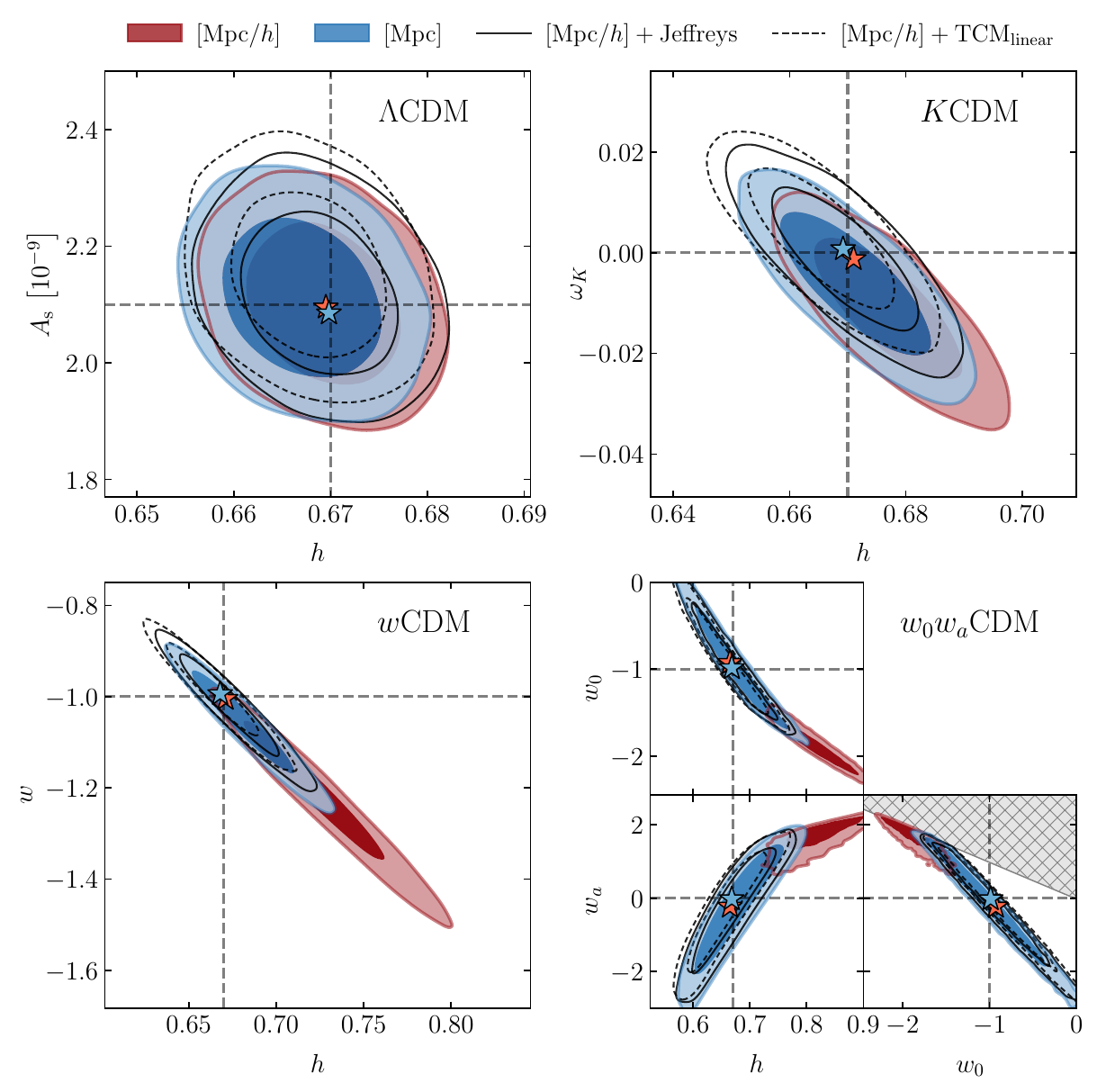}
    \caption{Marginalised posterior distributions of the Hubble expansion factor $h$ and selected cosmological parameters describing either the amplitude of primordial fluctuations, spatial curvature, or the evolution of the dark energy equation of state. The top panels show the results for \lcdm (left) and \kcdm (right), displaying constraints in the $(h,\As)$ and $(h,\omk)$ planes, respectively. The bottom panels present the corresponding constraints for \wcdm (left) and \wowacdm (right), focusing on the joint constraints of $h$ and the dark energy parameters, $(h,w)$ in the first case and $(h,\wo,\wa)$ in the second. The red filled contours represent the baseline configuration using $\Mpch$ units with standard priors. The dashed and solid black contours show alternative mitigation strategies applied to the $\Mpch$ parametrisation: Jeffreys priors for nuisance parameters and the reparametrisation described in \cite{TseCarMor2026} (denoted as ${\rm TCM_{linear}}$). The blue contours represent the results obtained in $\Mpc$ units with standard priors, without any correction scheme. Red and blue stars mark the \map point, obtained via direct likelihood minimisation, for the $\Mpch$ and $\Mpc$ vanilla cases, respectively.}
    \label{fig:comp_units_allcases_evo}
\end{figure*}

\subsection{Comparison between $\Mpch$- and $\Mpc$-defined priors}
\label{sec:comp_units_priors}

In Fig.~\ref{fig:comp_units_allcases_evo}, we show the marginalised posterior distributions in selected subspaces of the \lcdm (top left), \kcdm (top right), \wcdm (bottom left) and \wowacdm (bottom right) parameter spaces. In particular, we focus on constraints on the Hubble parameter $h$ and on additional evolution parameters, namely the amplitude of primordial fluctuations $\As$ in the \lcdm case, the curvature density parameter $\omk$ in the \kcdm case, and the parameters describing the dark energy equation of state---$w$ in the \wcdm case, and $(\wo,\wa)$ in the \wowacdm case.\footnote{The full marginalised posterior distributions, including all the remaining evolution and shape parameters, can be found in Appendix~\ref{app:full_contours}.} We compare three configurations in $\Mpch$ units---the baseline with standard priors (vanilla), and the additional configurations with either Jeffreys priors or the ${\rm TCM_{linear}}$ reparametrisation based on $\sigtwelve$\footnote{The use of the ${\rm TCM_{linear}}$ reparametrisation based on $\sigtwelve$ coupled to $\Mpch$ units is justified here after the considerations at the end of Sect.~\ref{sec:reparam_schemes}.}---with the baseline configuration obtained with $\Mpc$ units. This comparison reveals that the $\Mpc$ vanilla case, despite using no explicit correction scheme, achieves comparable posterior distributions to the corrected $\Mpch$ configurations, without the need of resorting to more sophisticated priors.

In the \lcdm case, a mild shift in the posterior is observed when changing the unit system, although the baseline configuration in $\Mpch$ units is already able to recover the fiducial values, indicating that \pve are limited in this scenario. The residual shift observed in $\Mpc$ units can be ascribed to leftover projection effects rather than to a genuine bias in the likelihood: the \map point, obtained by direct minimisation of the marginalised $\chi^2$, correctly converges to the fiducial position in both unit conventions. In the \kcdm case, instead, the baseline constraints obtained in $\Mpch$ units exhibit a more noticeable shift, reaching the $\sim1$–$2\sigma$ level. The most stringent test case is provided by the \wcdm and \wowacdm parameter spaces, where the dark energy equation of state is left free to vary: here the differences between the two unit conventions become substantial, with the posterior distribution in the baseline fit with $\Mpch$ units appearing strongly dominated by \pve, leading to biased constraints: as shown in Fig.~\ref{fig:laplace_term_hw0war}, this typically results in overestimated values of $h$ and underestimated values of $w(z)$ (through low $w$ in the \wcdm case, or low $\wo$ and high $\wa$ in the \wowacdm case).

In all cases, the use of $\Mpc$ units leads to posterior distributions that more faithfully recover the expected degeneracies among cosmological parameters, while remaining consistent with the fiducial \lcdm values. Any apparent degradation in constraining power observed in the extended parameter spaces (in particular \wowacdm) is therefore a consequence of removing these prior-induced distortions, rather than a genuine loss of information. Conversely, the tighter constraints obtained in $\Mpch$ units arise from an artificial weighting of specific regions of the parameter space, as already highlighted in Fig.~\ref{fig:laplace_term_hw0war}, and should not be interpreted as real improvements.

When adopting optimized mitigation strategies, such as the use of Jeffreys priors or the ${\rm TCM_{linear}}$ reparametrisation, the inferred posterior distribution of the cosmological parameters becomes much more consistent with the standard results obtained in $\Mpc$ units, independently of the unit system used in the fit. While some residual differences persist among these approaches, suggesting potential gains from improved reparametrisations or more suitable prior choices, these discrepancies are significantly smaller than those observed with the baseline priors in $\Mpch$ units.

\begin{figure}
    \includegraphics[width=\columnwidth]{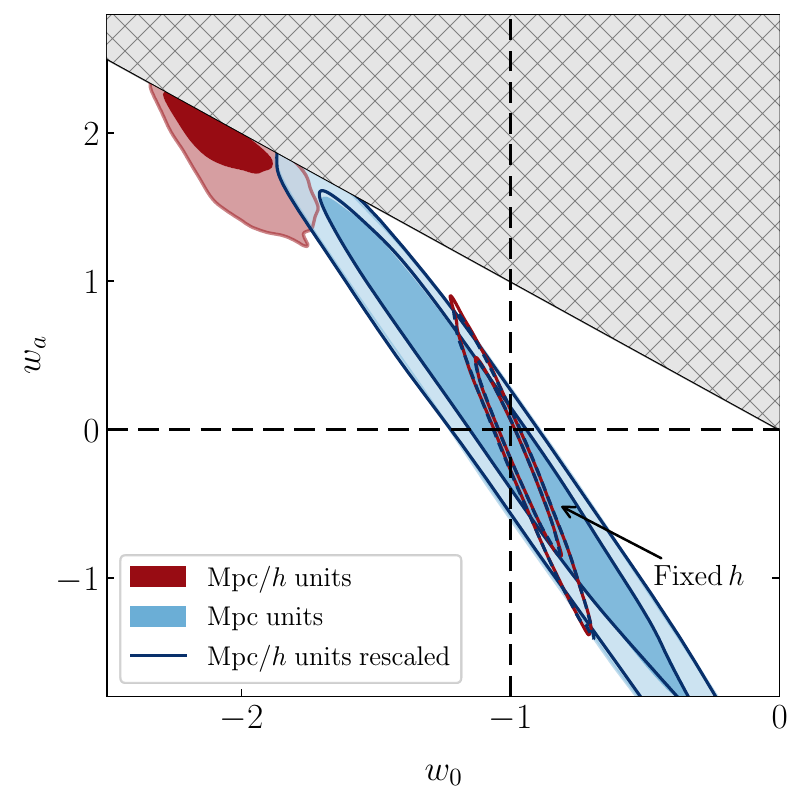}
    \caption{Comparison of the marginalised posterior distribution in the $(\wo, \wa)$ parameter space for different unit conventions. Filled red and filled light-blue contours correspond to the standard analyses in $\Mpch$ and $\Mpc$ units, respectively. The empty dark-blue contours show the $\Mpch$-rescaled case, obtained using the explicit $k/\knl$ reparametrisation of the counterterm and shot-noise contributions with $\knl$ and $\nbar$ fixed to the same values adopted in the native $\Mpc$ analysis, highlighting the agreement between the two when consistent priors are used. The thin solid red and dark-blue lines correspond to the standard $\Mpch$ and $\Mpc$ cases obtained when fixing $h$ to its fiducial value, illustrating that the residual discrepancy arises from sampling over $h$. The fiducial \lcdm point, with $h=0.67$, is indicated by the intersection of the dashed black lines. The hatched grey region above the threshold $\wo + \wa = 0$ is not explored by the likelihood.}
    \label{fig:w0wa_fixed-h}
\end{figure}

As a further test that the explicit $h$-dependence of the priors is the origin of these effects, we now fix the expansion rate to its fiducial \lcdm value and examine the resulting marginalised posterior distribution of $(\wo,\wa)$, as shown in Fig.~\ref{fig:w0wa_fixed-h}. As noted in the previous section, avoiding an explicit sampling of this parameter allows one to establish a one-to-one correspondence between parameter sets defined in $\Mpch$ and $\Mpc$ units—or, equivalently, between the non-linear scales $\knl$ entering the counterterm and shot-noise contributions. In this case, we find an almost perfect agreement between the two sets of contours, which are also consistently corrected for projection effects and recover the fiducial \lcdm values, with no significant residual bias. This provides direct evidence that $h$ is the parameter primarily responsible for the \pve discussed throughout this work: once $h$ is held fixed, the bias seen in the standard analysis disappears irrespective of the unit convention adopted. While small residual shifts may still arise from degeneracies with other evolution parameters, such as the scalar amplitude $\As$, fixing $h$ effectively captures the leading source of mismatch between the two parametrisations.\footnote{Fixing $h$ reduces the extent of the degeneracy between $\wo$ and $\wa$ compared to the case where $h$ is sampled over, since part of the elongation along this direction is itself induced by marginalising over $h$. A residual degeneracy nonetheless remains, reflecting the intrinsic geometrical degeneracy between $\wo$ and $\wa$ in the modelling of the dark energy equation of state.}

When $h$ is instead sampled over, the same marginalised posterior can be recovered in either unit system, provided that consistent priors are imposed in the two cases. In practice, this corresponds to defining priors on dimensional parameters independently of the unit convention adopted in the fit, or, equivalently, fixing the non-linear scale in a common unit system. This is illustrated in Fig.~\ref{fig:w0wa_fixed-h}, where the $\Mpch$-rescaled case is obtained by adopting the explicit $k/\knl$ reparametrisation of the counterterm and shot-noise contributions introduced in Sect.~\ref{sec:parametrisation}---so that the sampled amplitudes are dimensionless and equal in both unit systems---while fixing both $\knl$ and $\nbar$ to the same numerical values used in the native $\Mpc$ analysis, rather than rescaling them by the corresponding power of $h$ as would be customary when nominally working in $\Mpch$ units. It is precisely this choice---fixing the non-linear scale and number density in $\Mpc$ units---that yields a better-behaved parameter space and allows this case to reproduce results in excellent agreement with the native $\Mpc$ case. Equivalently, the same baseline $\Mpch$ contours could have been recovered by performing the fit in $\Mpc$ units while instead fixing $\knl$ and $\nbar$ to their $\Mpch$ values. In summary, it is not the unit system adopted to perform the fit that determines the resulting posterior, but rather the parameters on which the priors are actually imposed: whenever priors are placed on parameters that explicitly carry a factor of $h$, \pve are induced, regardless of the nominal unit convention used in the analysis.

\subsection{Impact from different reparametrisation schemes}
\label{sec:reparam_schemes}

\begin{figure*}
    \includegraphics[width=2\columnwidth]{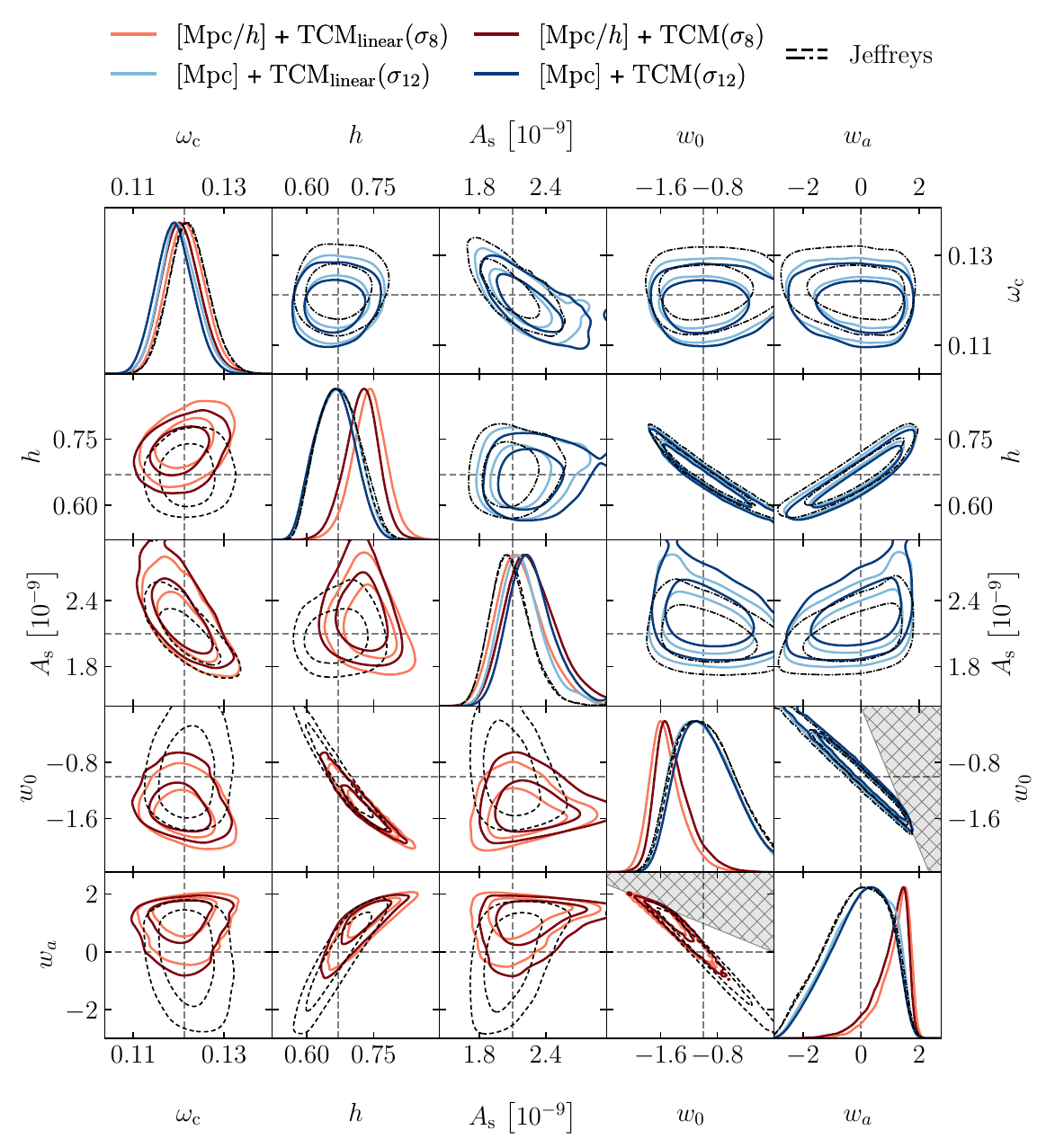}
    \caption{Comparison of the marginalised posterior distributions in the \wowacdm parameter space obtained with the linear (${\rm TCM_{linear}}$, light shades) and full (TCM, dark shades) implementations of the TCM reparametrisation scheme of \cite{TseCarMor2026}. The lower-left triangle (red) shows results obtained in $\Mpch$ units, with the reparametrisation calibrated on $\sigma_8$, while the upper-right triangle (blue, mirrored) shows the corresponding results in $\Mpc$ units, calibrated on $\sigtwelve$. In each triangle, the black dashed and dot-dashed contours show the results obtained when imposing Jeffreys priors on the linear nuisance parameters, in $\Mpch$ and $\Mpc$ units, respectively. The hatched region indicates $\wo+\wa>0$, excluded by the likelihood.}
    \label{fig:triangle_mirrored_TCM}
\end{figure*}

Having shown that working in $\Mpc$ units is sufficient to mitigate \pve regardless of the specific treatment of the remaining nuisance parameters, we now examine the performance of an explicit reparametrisation scheme, focusing on the TCM reparametrisation of \cite{TseCarMor2026}. We compare its linear (${\rm TCM_{linear}}$) and full (TCM) implementations in both unit conventions, each calibrated in terms of the amplitude parameter customarily associated with the corresponding unit convention: $\sigma_8$ in $\Mpch$ units and $\sigtwelve$ in $\Mpc$ units. In the latter case, the choice of $\sigtwelve$ is strictly necessary, as using $\sigma_8$ would introduce an explicit $h$-dependence into the reparametrisation. We then assess the impact of the amplitude parameter itself, by replacing $\sigma_8$ with $\sigtwelve$ in the ${\rm TCM_{linear}}$ reparametrisation applied in $\Mpch$ units. In all cases, we compare the reparametrisation-based results against those obtained with Jeffreys priors, which we take as the reference case that the reparametrisation-based approaches are expected to reproduce.

Figure~\ref{fig:triangle_mirrored_TCM} shows the results of the first comparison, with the $\Mpch$ and $\Mpc$ cases displayed in the lower-left and upper-right (mirrored) panels, respectively. We first focus on the $\Mpc$ case, where both the ${\rm TCM_{linear}}$ and TCM reparametrisations converge onto the same posterior distribution as that obtained with Jeffreys priors (black dot-dashed contours), indicating excellent agreement between the two mitigation strategies. Nevertheless, a small residual discrepancy persists between the TCM and ${\rm TCM_{linear}}$ posteriors, most noticeably in the marginalised constraints on $\As$. This could be ascribed to the different geometry of the parameter space induced by the two parametrisations, as well as to the different prior widths adopted in each case: in the full TCM implementation, the prior on nuisance parameters that enter the model non-linearly is additionally rescaled to account for the expectation value of the corresponding parameter (e.g. the prior of $\bone$ is rescaled to be consistent with $\bone\sqrt{A_{\rm AP}}\,\sigtwelve$), a rescaling that is not applied in the ${\rm TCM_{linear}}$ case. We explicitly tested which of the reparametrised nuisance parameters is responsible for this shift, finding that it is driven by the reparametrisation of the quadratic and tidal bias parameters, $\btwo$ and $\gtwo$, rather than by that of the linear bias $\bone$, which proves comparatively safe. This can be traced back to the correspondingly broad, flat priors imposed on $\btwo$ and $\gtwo$: replacing them with narrower, localised Gaussian priors improves the agreement between the two implementations. These differences remain minimal, however, and we take the overall agreement between the two implementations as evidence of the robustness of the reparametrisation approach, which is most effective when applied in $\Mpc$ units.

\begin{figure}
    \includegraphics[width=\columnwidth]{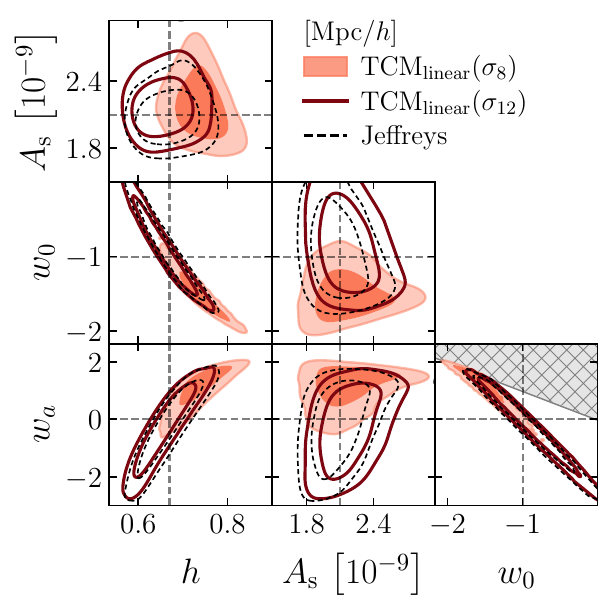}
    \caption{Marginalised posterior distributions in the $(h,\As,\wo,\wa)$ subspace of the \wowacdm parameter space obtained in $\Mpch$ units with the ${\rm TCM_{linear}}$ reparametrisation of \cite{TseCarMor2026}, calibrated either on $\sigma_8$ (filled light-red contours) or on $\sigtwelve$ (empty dark-red contours). The black dashed contours show the results obtained when imposing Jeffreys priors on the linear nuisance parameters. Dashed grey lines indicate the fiducial \lcdm values, while the hatched region indicates $\wo+\wa>0$, excluded by the likelihood.}
    \label{fig:TCMlinear_sigma_hMpc}
\end{figure}

The picture is different in $\Mpch$ units: while the $(\wo,\wa)$ posterior obtained with the $\sigma_8$-calibrated reparametrisation lies along the same degeneracy direction explored in the Jeffreys case (black dashed contours), the agreement holds only along one side of this degeneracy, the upper-left region of the $\wo$--$\wa$ panel, and the posterior of $h$ is correspondingly shifted towards higher values. We find that this limitation is largely driven by the choice of $\sigma_8$ as the amplitude parameter, as shown in Fig.~\ref{fig:TCMlinear_sigma_hMpc}, which compares the ${\rm TCM_{linear}}$ results obtained in $\Mpch$ units when calibrating the reparametrisation on either $\sigma_8$ or $\sigtwelve$. Indeed, replacing $\sigma_8$ with $\sigtwelve$ removes this discrepancy, recovering the full extent of the $(h,\wo,\wa)$ degeneracy and yielding a posterior in close agreement with the Jeffreys case and with the fiducial cosmology. Notably, the resulting constraints on $\As$ are also fully consistent with the fiducial value. This is particularly relevant since alternative implementations of the TCM scheme that do not rescale the nuisance parameters by any amplitude parameter tracking the growth of density fluctuations could lead to a systematic shift in $\As$. Our results therefore indicate that the TCM reparametrisation can also be an effective strategy to mitigate \pve when working in $\Mpch$ units, and that adopting $\sigtwelve$ as the amplitude parameter yields better performance than $\sigma_8$.\footnote{This can be understood by noting that the radius $R=8\,\Mpch$ defining $\sigma_8$ corresponds to a physical scale that depends on the value of $h$ \citep{Sanchez2020, SanRuiJar2022, SemSanPez2022, ForFavGom2025}. When $h$ is poorly constrained, as is the case when exploring extended dark-energy models, $\sigma_8$ is therefore obtained by integrating the linear power spectrum over substantially different ranges of scales at different points of the parameter space (see Fig.~2 of \cite{SemSanPez2022}), making it a poor proxy for the amplitude of the power spectrum. By contrast, $\sigtwelve$ is always defined on the same scale, $R=12\,\Mpc$, independently of the value of $h$, and therefore provides a consistent measure of the clustering amplitude across the whole parameter space.} A more robust assessment of this configuration would nonetheless require dedicated tests based on synthetic data sets tailored to the specific catalogue to be analysed.

As a further consistency check, in Appendix~\ref{app:withAP} we repeat some of the tests run in this section using synthetic data vectors that include Alcock-Paczynski distortions due to a wrong choice of fiducial cosmology, for two different dark-energy equations of state (one broadly consistent with recent DESI constraints \citep{AbdAguAhl2025, DESI_FS}), showing consistent results in both cases.

\begin{figure*}
    \includegraphics[width=2\columnwidth]{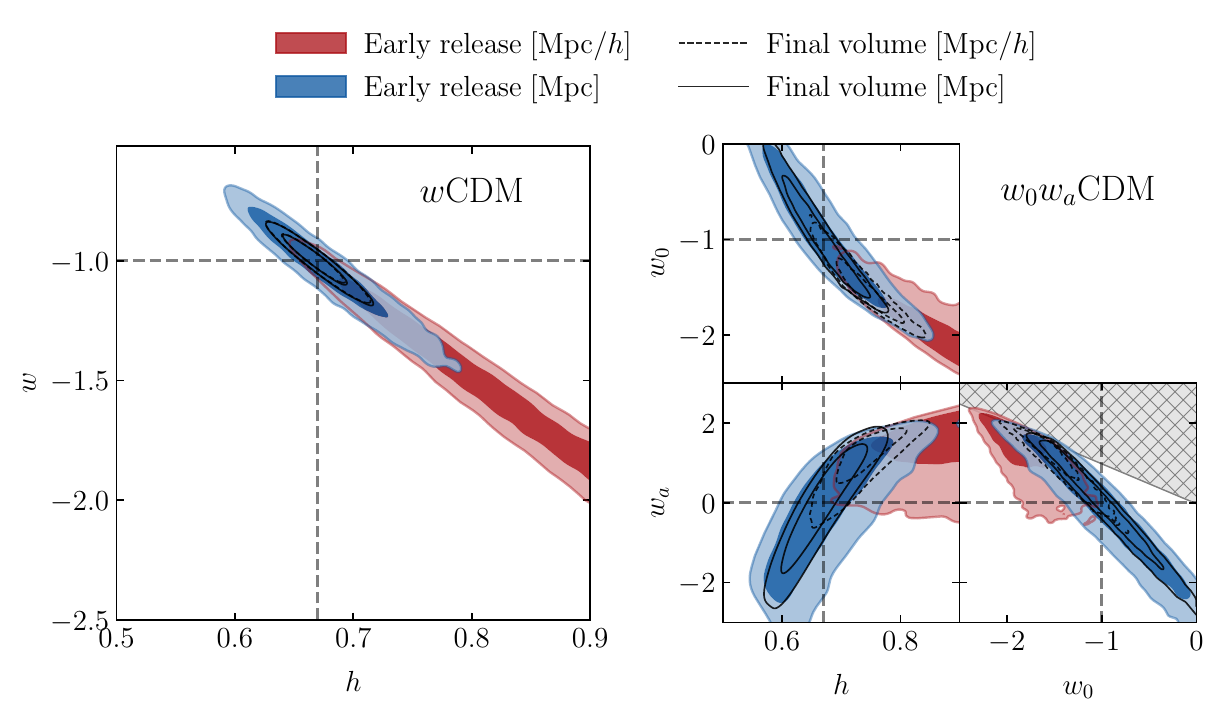}
    \caption{Comparison of the marginalised posterior distributions in the \wcdm (left) and \wowacdm (right) parameter spaces, obtained with the ${\rm TCM_{linear}}(\sigma_8)$ reparametrisation in $\Mpch$ units and the ${\rm TCM_{linear}}(\sigtwelve)$ reparametrisation in $\Mpc$ units, for two survey volumes of different statistical power. Filled contours, coloured by unit ($\Mpch$ in red, $\Mpc$ in blue), show the results obtained for the reduced-area case of $2\,500 \, {\rm deg}^2$; black contours (dashed for $\Mpch$ and solid for $\Mpc$) show the corresponding results for the fiducial, full survey volume ($15\,000 \, {\rm deg}^2$) used as the reference throughout this work. The hatched region indicates $\wo+\wa>0$, excluded by the likelihood.}
    \label{fig:comp_DR1_vs_DR3}
\end{figure*}

\subsection{Robustness to the statistical power of the data}
\label{sec:comp_DR1_vs_DR3}

To assess the robustness of our conclusions to the statistical power of the data, we repeat the analysis using synthetic data vectors and covariances representative of an idealised survey with the same number density (and hence the same per-mode signal-to-noise) as our fiducial setup, but covering a reduced effective sky area of $\sim2\,500\,{\rm deg}^2$, rather than the $\sim15\,000\,{\rm deg}^2$ assumed elsewhere in this work. These values are not meant to be exact, but broadly representative of the early and full-mission data volumes expected for the \Euclid survey.

Figure~\ref{fig:comp_DR1_vs_DR3} compares the marginalised posterior distributions obtained with the ${\rm TCM_{linear}}$ reparametrisation, in $\Mpch$ and $\Mpc$ units, for the two survey volumes described above. The same conclusions drawn for the fiducial, full survey volume also hold for the reduced-area case, and are in fact exacerbated by its lower statistical power. In $\Mpch$ units, the $\sigma_8$-calibrated reparametrisation is unable to fully correct for \pve: in the \wcdm case, the reduced-area posterior remains substantially shifted from the fiducial value, similarly to what is found for the non-reparametrised, baseline case at full survey volume; in the \wowacdm case, the resulting posterior instead exhibits a break along the $(h,\wo,\wa)$ degeneracy direction, losing part of the extended degeneracy recovered in $\Mpc$ units, and for the reduced-area case this translates into a marginalised posterior biased by more than $2\sigma$ from the fiducial \lcdm values. By contrast, the $\sigtwelve$-calibrated reparametrisation in $\Mpc$ units keeps the contours well centred on the fiducial values for both survey volumes, with the lower statistical power of the reduced-area case reflected only in a broadening of the contours, rather than in any additional bias.

This test is especially relevant for the upcoming analysis of \Euclid DR1 data, whose constraining power is expected to be of the same order as the reduced-area case considered here, further compounded by systematic effects that reduce both the effective number density and the observable sky area. In App.~\ref{app:desi_like} we show that these findings can be partially modified when using an alternative synthetic data set that also includes lower-redshift samples, as in \desi, where the evolution of the dark energy equation of state can be more easily constrained; nevertheless, even in that case, the improved performance of the $\Mpc$+${\rm TCM_{linear}}(\sigtwelve)$ reparametrisation is never surpassed.

\subsection{Robustness to inclusion of \bao constraints}

\begin{figure*}
    \includegraphics[width=2\columnwidth]{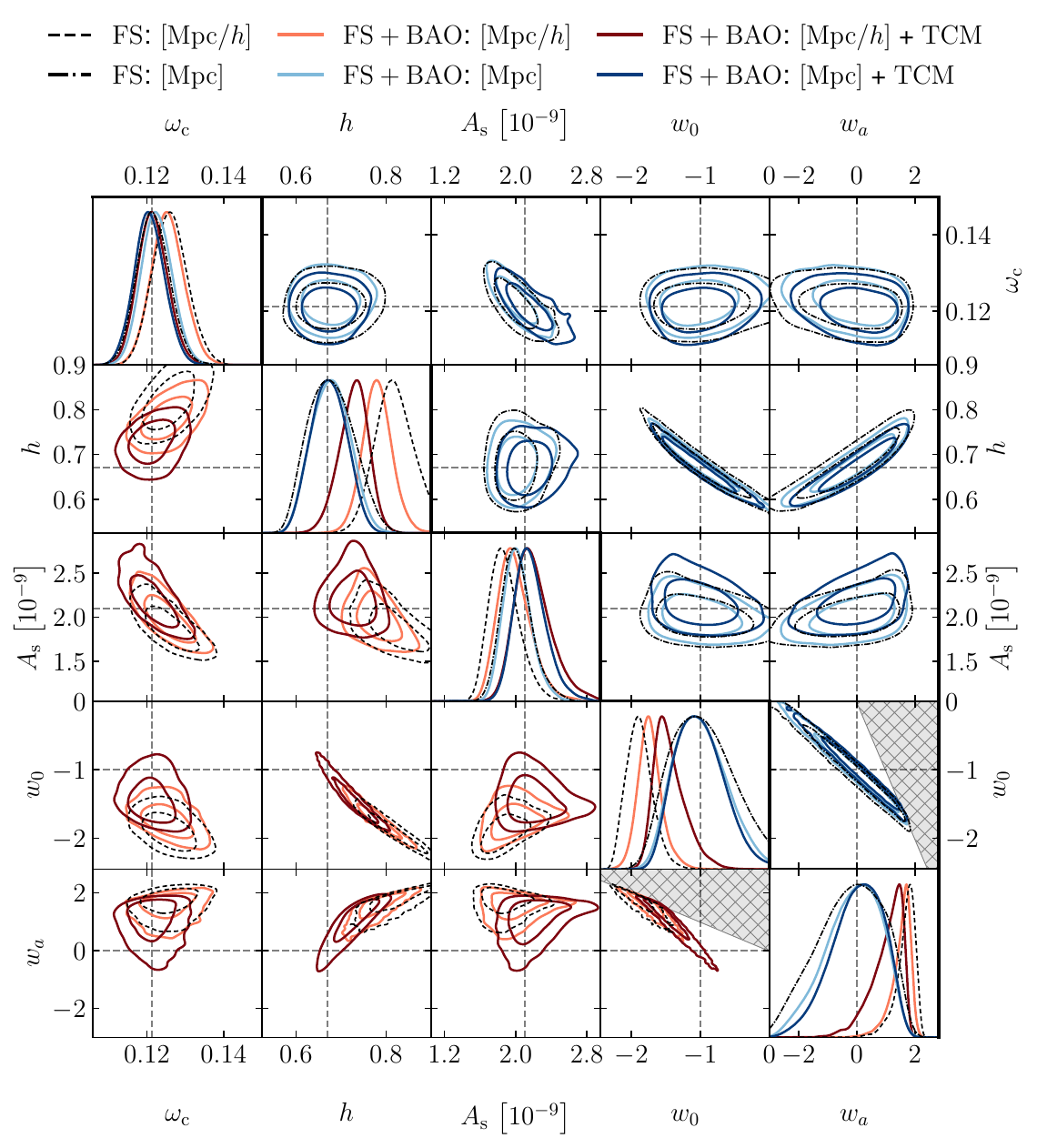}
    \caption{Comparison of the marginalised posterior distributions in the \wowacdm parameter space, illustrating the impact of combining full-shape (FS) information with post-reconstruction \bao\ measurements (FS+BAO). The lower-left triangle shows results obtained in $\Mpch$ units, while the upper-right triangle (mirrored) shows the corresponding results in $\Mpc$ units. In each panel, the black dashed ($\Mpch$) and dot-dashed ($\Mpc$) contours show the FS-only baseline; the light-shaded contours show the standard FS+BAO case; and the dark-shaded contours show the FS+BAO case after applying the ${\rm TCM_{linear}}$ reparametrisation, calibrated on $\sigma_8$ in $\Mpch$ units and on $\sigtwelve$ in $\Mpc$ units. The hatched region indicates $\wo+\wa>0$, excluded by the likelihood.}
    \label{fig:triangle_mirrored_FSBAO}
\end{figure*}

The \fs information considered so far---i.e., the power spectrum multipoles used throughout this work---is routinely combined, in realistic analyses and in particular for all Stage~III and Stage~IV spectroscopic surveys, with the geometrical information extracted from the post-reconstructed \bao\ signal \citep{EisSeoSir2007, AlaAtaBai2017, PhiIvaSim2020, AlaAubAvi2021, DESI2024}. The \bao\ likelihood provides comparatively tight constraints on the transverse and line-of-sight dilation parameters, $\alpha_\perp$ and $\alpha_\parallel$, which in turn tighten the constraints on the parameters governing the expansion history, such as $h$, $\wo$, and $\wa$. It is therefore worth testing whether the inclusion of this information alters the impact of \pve, and the effectiveness of the reparametrisation strategies discussed above, when combined with the \fs likelihood.

The synthetic \bao\ data are generated assuming the fiducial cosmology to be correct, such that the \bao\ shift parameters $\alpha_\perp$ and $\alpha_\parallel$ are set to unity by default. The covariance between $\alpha_\perp$ and $\alpha_\parallel$ is estimated by inverting the Fisher matrix of the reconstructed power spectrum model, marginalising over all nuisance parameters \citep{BeuSeoRos2016}. The cross-covariance between the pre-reconstruction power spectrum multipoles and the \bao\ parameters is then obtained by propagating the Fisher matrix covariance through the derivatives of the pre-reconstruction model with respect to the \bao\ model parameters; the details of this model are not reported here, as they are not the focus of this work.

Figure~\ref{fig:triangle_mirrored_FSBAO} shows the impact of adding post-reconstruction \bao\ information to the FS-only likelihood, in the \wowacdm parameter space, comparing the $\Mpch$ (lower-left panels) and $\Mpc$ (upper-right panels, mirrored) unit conventions. In $\Mpch$ units, the addition of \bao\ information tightens the constraints on the $\{h,\wo,\wa\}$ triplet, reflecting the additional geometrical information carried by the \bao\ dilation parameters, $\alpha_\perp$ and $\alpha_\parallel$. However, even after applying the ${\rm TCM_{linear}}$ reparametrisation to this FS+BAO combination, the resulting posterior remains shifted from the true \lcdm values by up to $\sim1\sigma$, for this particular configuration: the residual \pve is not removed by the reparametrisation, closely mirroring the behaviour already observed in the FS-only case.

By contrast, in $\Mpc$ units the contours remain close to the fiducial values throughout. Adding \bao\ information to the FS-only likelihood tightens the constraints, as expected from the inclusion of genuinely new data; applying the ${\rm TCM_{linear}}$ reparametrisation based on $\sigtwelve$ to this same FS+BAO combination further sharpens the contours without introducing any additional drift. This indicates that, once \pve are mitigated by adopting $\Mpc$ units, both genuinely new data (\bao) and reparametrisation-driven prior corrections behave as expected, tightening the constraints while preserving their accuracy. In $\Mpch$ units, instead, the same reparametrisation is unable to fully remove the residual bias induced by the prior, even though it still benefits from the additional constraining power brought by the \bao\ data.

\subsection{Comparison between different \texorpdfstring{\rsd}{RSD} models}

As a final check, we test whether the conclusions reached so far are robust against the specific choice of \rsd model adopted to describe the galaxy power spectrum, comparing the \eft and \vdg frameworks introduced in Sect.~\ref{sec:parametrisation}. For this purpose, we additionally generate synthetic data vectors for the \vdg model, following the same procedure adopted for the \eft data vectors used throughout the rest of this work.

\begin{figure*}
    \includegraphics[width=2\columnwidth]{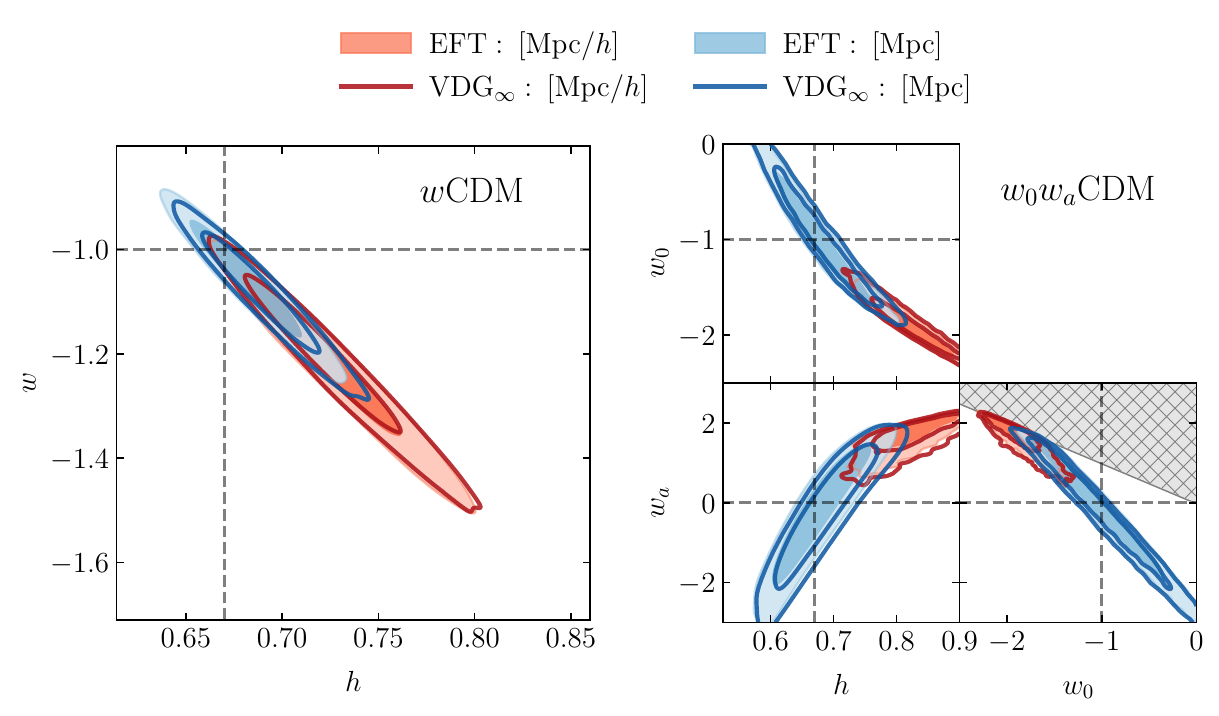}
    \caption{Same as in the bottom panels of Fig.~\ref{fig:comp_units_allcases_evo}, but showing a comparison between the performance of the \eft and \vdg models in terms of \pve, when working in either $\Mpch$ or $\Mpc$ units.}
    \label{fig:comp_units_models_evo_wCDM_w0waCDM}
\end{figure*}

In Fig.~\ref{fig:comp_units_models_evo_wCDM_w0waCDM} we show a comparison between the \eft and \vdg modelling approaches in terms of \pve, for analyses performed in either $\Mpch$ or $\Mpc$ units. When working in $\Mpch$ units, both models yield broadly consistent constraints, with only minor residual differences in the posterior shapes that are completely subdominant with respect to the main projection effects. In a similar way we observe a consistent agreement between the two models when imposing priors in $\Mpc$ units. In this case, the impact of \pve is significantly reduced for both models, and the posteriors suggest that they share comparable performances.\footnote{We note that this does not provide information on the robustness of the individual models or their range of validity, as the data vectors considered here are synthetic data generated using the same theoretical model adopted in the likelihood exploration.}

This behaviour can be understood in terms of the structure of the two modelling frameworks. Both the \eft and \vdg models share the same number of free parameters and a largely one-to-one correspondence between them, differing only in a small subset of parameters controlling the velocity dispersion (namely $\avir$ and $\cnlo$). As a consequence, both models exhibit a comparable sensitivity to the emergence of \pve, which explains the consistent behaviour observed in the figure.

\section{Conclusions}
\label{sec:conclusions}

In this paper, we have investigated the dependence of prior-volume effects on the choice of units used to express two-point clustering measurements, using as observables the Legendre multipoles of the galaxy power spectrum. These effects naturally arise in a Bayesian inference framework when the constraining power of the data is limited, and the effective hypervolume associated with the marginalised nuisance parameters varies significantly across the parameter space of interest, namely the cosmological parameters. This is typically the case for galaxy clustering analyses based on state-of-the-art models of non-linear evolution, galaxy bias, and \rsd, such as the \eft and \vdg models.

The outcome of the analysis is that the Laplace term arising in the marginalised $\chi^2$, which encodes the effect of the effective hypervolume of the nuisance parameters as a function of the cosmological ones, exhibits a reduced sensitivity to variations in the cosmological parameters when the fit is performed in $\Mpc$ units as opposed to the more commonly used $\Mpch$ convention. More specifically, iso-contours of the Laplace contribution indicate that \pve tend to favour lower marginalised values of the dark energy equation of state (parametrised here by $w$ or by $(\wo, \wa)$ in the \wcdm and \wowacdm parameter spaces, respectively) and higher values of the Hubble expansion rate $h$, as indicated in recent analyses of BOSS \citep{CarMorPou2023} and DESI \citep{DESI_FS}. A qualitatively similar behaviour is observed when extending the analysis to models with non-zero spatial curvature. We further confirm that $h$ itself, rather than the choice of units per se, is the parameter primarily responsible for the emergence of these effects: fixing $h$ to its fiducial value removes the residual discrepancy between the $\Mpch$ and $\Mpc$ parametrisations, irrespective of the unit convention adopted.

Extending the analysis to the full Bayesian inference of cosmological parameters, we find that adopting $\Mpc$ units alone, without any additional correction, is sufficient to recover posterior distributions that are largely free from prior-volume induced shifts, across all the cosmological models considered (\lcdm, \kcdm, \wcdm, and \wowacdm). These results are obtained assuming an idealised, full-mission Stage-IV survey configuration representative of \Euclid, with four redshift bins spanning $0.9<z<1.8$ and a Gaussian covariance matrix representative of the corresponding survey volume. In the most affected cases, namely \wcdm and \wowacdm, this simple change of unit convention reduces discrepancies of several standard deviations, typically found when adopting the traditional $\Mpch$ priors, down to a level fully consistent with the fiducial cosmology. The performance of this approach is comparable to that of more elaborate mitigation strategies applied within the $\Mpch$ parametrisation, such as Jeffreys priors for the linearly-entering nuisance parameters or the reparametrisation of \cite{TseCarMor2026} in terms of the \apdist amplitude and $\sigtwelve$, while requiring no additional modelling effort.

We further verified that, once \pve are mitigated by working in $\Mpc$ units, the residual sensitivity to the specific implementation of the reparametrisation strategy is strongly suppressed: the linear and full TCM implementations, both based on $\sigtwelve$, yield consistent constraints that closely track the Jeffreys reference case, with the full implementation only leading to slightly higher values of $\As$. This indicates that the robustness of this mitigation approach is controlled more substantially by the units in which priors are imposed and the non-linear scale and number density are fixed than by the specific details of the reparametrisation scheme (see Sect.~\ref{sec:parametrisation}). The picture is different in $\Mpch$ units with the reparametrisation based on $\sigma_8$: here the reparametrisation only partially corrects for \pve, reproducing the Jeffreys posterior along only part of the $(\wo,\wa)$ degeneracy direction. However, we find that this limitation is largely driven by the choice of $\sigma_8$ as the amplitude parameter: when the TCM reparametrisation is instead based on $\sigtwelve$, the analysis in $\Mpch$ units also recovers a posterior in close agreement with the Jeffreys case and with the fiducial cosmology, including for $\As$. This suggests that pipelines natively working in $\Mpch$ units may also achieve a substantial reduction of \pve by adopting $\sigtwelve$ instead of $\sigma_8$ in the TCM reparametrisation, although a more robust assessment of this configuration would require dedicated tests tailored to the data set to be analysed. Similarly, the reduction of \pve in $\Mpc$ units is robust against the specific \rsd model adopted, with the \eft and \vdg frameworks exhibiting closely comparable behaviour, consistent with their similar parameter structure.

\begin{table*}
  \caption{Fiducial values of the nuisance parameters adopted to generate the synthetic \eft (left) and \vdg (right) data vectors, for each redshift bin, labelled by its effective redshift $z_{\rm eff}$. Dimensional parameters are expressed in $\Mpch$ units. In both cases, the scale-dependent shot-noise parameters $\nptwozero$ and $\nptwotwo$ are set to zero, as is $\cnlo$ for the \eft model.}
  \vspace{6pt}
  \centering
  \renewcommand{\arraystretch}{1.3}
  \setlength{\tabcolsep}{10pt}
  \begin{tabular}{|c|cccc|cccc|}
    \hline
    & \multicolumn{4}{c|}{\eft} & \multicolumn{4}{c|}{\vdg} \\
    \hline
    $z_{\rm eff}$ & 1.0 & 1.2 & 1.4 & 1.65 & 1.0 & 1.2 & 1.4 & 1.65 \\
    \hline
    $\bone$ & 1.412 & 1.769 & 2.039 & 2.496 & 1.393 & 1.755 & 2.031 & 2.506 \\
    $\btwo$ & 0.698 & 0.870 & 1.161 & 2.010 & $-0.417$ & $-0.353$ & $-0.090$ & 0.635 \\
    $\gtwo$ & $-0.156$ & $-0.300$ & $-0.400$ & $-0.555$ & $-0.149$ & $-0.294$ & $-0.397$ & $-0.558$ \\
    $\gtwoone$ & 0.095 & $-0.184$ & $-0.244$ & $-0.333$ & $-0.090$ & $-0.180$ & $-0.242$ & $-0.335$ \\
    $\czero\,[(\Mpch)^2]$ & 13.662 & 16.635 & 16.629 & 24.197 & $-0.128$ & $-1.816$ & $-3.303$ & $-8.693$ \\
    $\ctwo\,[(\Mpch)^2]$ & 20.171 & 23.759 & 22.164 & 27.659 & $-3.737$ & $-6.963$ & $-11.779$ & $-13.918$ \\
    $\cfour\,[(\Mpch)^2]$ & 4.351 & 4.645 & 3.963 & 3.972 & $-1.276$ & $-1.862$ & $-2.936$ & $-3.579$ \\
    $\avir\,[\Mpch]$ & $--$ & $--$ & $--$ & $--$ & 2.291 & 2.385 & 2.571 & 2.225 \\
    $\npzero$ & 0.056 & 0.151 & 0.144 & 0.309 & 0.375 & 0.231 & 0.148 & 0.194 \\
    \hline
  \end{tabular}
  \label{tab:fiducial_nuisance}
\end{table*}

We tested the robustness of these conclusions against the statistical power of the data, repeating the analysis with a reduced effective sky area of $\sim2\,500\,{\rm deg}^2$, representative of an early data release, as opposed to the $\sim15\,000\,{\rm deg}^2$ full-mission volume assumed elsewhere in this work. The same conclusions hold in this reduced-constraining-power setting, and are in fact exacerbated: in \wcdm, the $\Mpch$\,+\,${\rm TCM}(\sigma_8)$ posterior remains substantially displaced from the fiducial value, while in \wowacdm it remains biased by more than $2\sigma$; the $\Mpc$+${\rm TCM}(\sigtwelve)$ combination, by contrast, remains well centred on the fiducial cosmology throughout, with the reduced statistical power reflected only in a broadening of the contours. This result is especially relevant for the upcoming analysis of \Euclid DR1 data, expected to have a constraining power of this order.

We also examined the impact of combining our \fs likelihood with post-reconstruction \bao\ measurements. Including \bao\ information tightens the constraints in both unit systems, but does not alter the overall picture: in $\Mpch$ units, the ${\rm TCM}(\sigma_8)$ reparametrisation remains unable to fully remove the residual bias induced by the prior, even after combining with \bao, while in $\Mpc$ units the ${\rm TCM}(\sigtwelve)$ combination continues to yield unbiased, tightened constraints.

Finally, in two dedicated appendices we further stress-tested the robustness of the $\Mpc$+${\rm TCM}(\sigtwelve)$ mitigation strategy under more realistic conditions. In Appendix~\ref{app:desi_like}, using a DESI-like synthetic data set, we show that the performance of the $\Mpch$+${\rm TCM}(\sigma_8)$ combination depends significantly on which redshift bins are included in the analysis: the reparametrisation is far more effective when only low-redshift tracers are considered, where the sensitivity to the dark-energy equation of state is stronger, than when high-redshift tracers dominate the fit. The $\Mpc$+${\rm TCM}(\sigtwelve)$ combination, by contrast, performs consistently well regardless of the redshift bins used. In Appendix~\ref{app:withAP}, we further show that this conclusion is unchanged when Alcock-Paczynski distortions are already present in the data---a more realistic setting, since the true cosmology is unknown when analysing real data---with the $\Mpc$+${\rm TCM}(\sigtwelve)$ combination remaining the best-performing configuration for two different dark-energy equations of state.

These results indicate that the choice of physical units adopted to express nuisance parameters---rather than the specific prior size, reparametrisation, or modelling choices---is the dominant driver of \pve in full-shape galaxy clustering analyses. Adopting $\Mpc$ units therefore provides a simple, physically motivated strategy to mitigate this systematic, with direct relevance for the forthcoming generation of Stage-IV spectroscopic surveys, including \Euclid and DESI.

\section*{Acknowledgements}

The authors thank Davide Bianchi, Pedro Carrilho, Emanuele Castorina, Martin Kärcher, Marco Marinucci, Chiara Moretti, Gabriele Parimbelli, Román Scoccimarro, Maria Tsedrik, and Pierre Zhang for useful discussions. The majority of the analysis has been run at the HPC of the Max-Planck Institute for Extraterrestrial Physics (MPE) in Garching. AP acknowledges support from the INAF AstroFit fellowship (Iniziative di Ricerca Fondamentale 2021-2025). AGS is supported by the Deutsche Forschungsgemeinschaft (DFG, German Research Foundation) under Germany's Excellence Strategy – EXC 2094 – 390783311. AE is funded by the Deutsche Forschungsgemeinschaft (DFG, German Research Foundation) under Germany’s Excellence Strategy EXC 3037 – 533607693– Unser dynamisches Universum. MC and GG acknowledge support from the Spanish Ministerio de Ciencia, Innovación y Universidades, project PID2021-128989NB. Our research made use of \texttt{matplotlib}, a \python \ library for publication quality graphics \citep{Hun2007}.

\appendix

\section{Fiducial values of the nuisance parameters}
\label{app:fiducial_nuisance}

Table~\ref{tab:fiducial_nuisance} lists the values of the nuisance parameters adopted to generate the synthetic \eft and \vdg data vectors in each of the four redshift bins. Dimensional parameters are expressed in $\Mpch$ units; the corresponding values in $\Mpc$ units are obtained by rescaling them by the appropriate power of $\hfid$, following Table~\ref{tab:unit_conversion}.

\section{Full marginalised contours of traditional \texorpdfstring{\lcdm}{ΛCDM} extensions and impact of $A_{\rm AP}$ on shot-noise terms}
\label{app:full_contours}

\begin{figure*}
    \includegraphics[width=2\columnwidth]{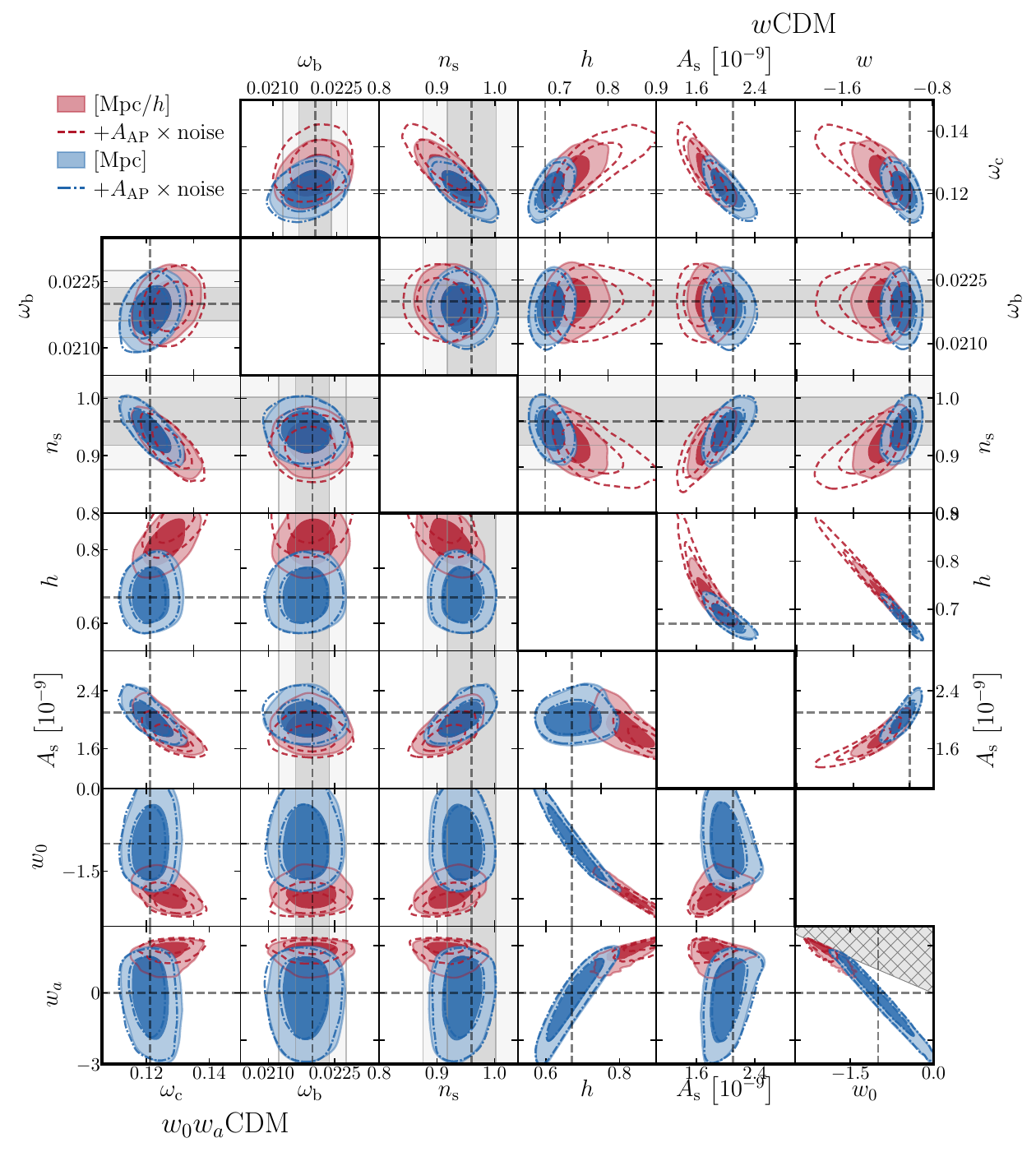}
    \caption{Same as in the bottom panels of Fig.~\ref{fig:comp_units_allcases_evo}, but additionally showing the remaining cosmological parameters, including the shape parameters $(\omc, \omb, \ns)$ and the evolution parameter $\As$, sampled when exploring the \wcdm (bottom left) and \wowacdm (top right) parameter spaces. The comparison is performed using only the baseline priors in either $\Mpch$ or $\Mpc$ units. Dashed and dot-dashed contours additionally show the effect of rescaling the stochastic contribution by $A_{\rm AP}$. We impose a BBN prior on $\omb$ and a $10\times$ Planck prior on $\ns$, as highlighted by the shaded grey bands in the corresponding panels.}
    \label{fig:comp_units_allcases_wCDM_w0waCDM}
\end{figure*}

In this appendix, we present the full marginalised posterior distributions obtained for the \wcdm and \wowacdm models considered in Sect.~\ref{sec:results}, complementing the subset of evolution parameters shown in Fig.~\ref{fig:comp_units_allcases_evo} with the remaining shape parameters $(\omc,\omb,\ns)$, shown in Fig.~\ref{fig:comp_units_allcases_wCDM_w0waCDM}. The corresponding results for the \lcdm and \kcdm parameter spaces follow the same trend already discussed for the evolution-parameter subset in Sect.~\ref{sec:results}---switching from $\Mpch$ to $\Mpc$ units restores agreement with the fiducial cosmology across all sampled parameters, including the shape parameters $(\omc,\omb,\ns)$---and are therefore not shown here.

In Fig.~\ref{fig:comp_units_allcases_wCDM_w0waCDM} we show the full two-dimensional marginalised posterior distributions obtained in the \wcdm and \wowacdm parameter spaces, using either the baseline set of priors in $\Mpch$ or $\Mpc$ units. As discussed in Sect.~\ref{sec:results}, the degeneracy between the expansion rate $h$ and the dark-energy parameters, $w$ or $(\wo,\wa)$, is properly recovered when switching from $\Mpch$ to $\Mpc$ units. This correction propagates naturally to the other cosmological parameters sampled in the analysis, starting with the scalar amplitude $\As$. Since $\As$ controls the overall amplitude of density fluctuations, it is also degenerate with other evolution parameters. Its apparent underestimation---one of the clearest indications of \pve---is mitigated when using $\Mpc$ units.

Turning to the shape parameters, we note that the marginalised posterior of $\omc$ is recovered with remarkable accuracy in $\Mpc$ units, even in the \wowacdm case where \pve are most severe. This indicates that the benefit of adopting $\Mpc$ units extends beyond the evolution sector directly entangled with the dark-energy equation of state, restoring unbiased constraints on the shape of the matter power spectrum as well. The same applies to $\omb$ and $\ns$, which, despite being sampled with Gaussian priors centred on their fiducial values, can still be affected by \pve. This is particularly relevant for the scalar spectral index, which, when estimated using the $\Mpch$ parametrisation, typically shifts towards values lower than the fiducial one.

\begin{figure}
    \includegraphics[width=\columnwidth]{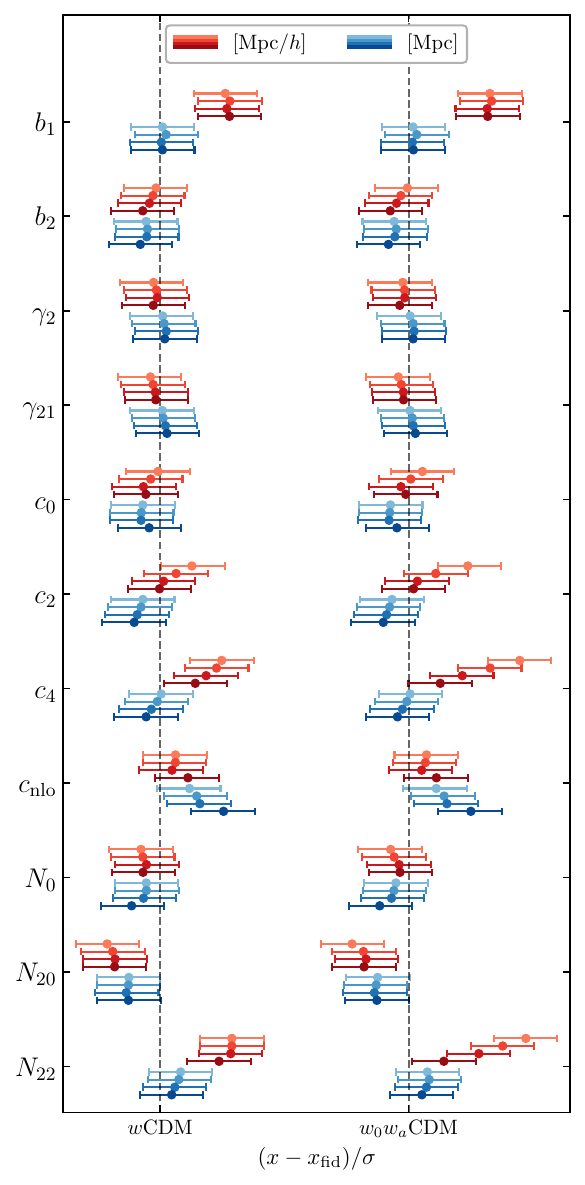}
    \caption{Weighted posterior mean of all the nuisance parameters of the model---the linear and quadratic bias $\bone$ and $\btwo$, the tidal bias parameters $\gtwo$ and $\gtwoone$, the counterterms $\czero$, $\ctwo$, and $\cfour$, the next-to-leading-order counterterm $\cnlo$, and the shot-noise parameters $\npzero$, $\nptwozero$, and $\nptwotwo$---for each of the four redshift bins, expressed as a deviation from the fiducial value in units of the corresponding posterior $1\sigma$ width, comparing the \wcdm (left) and \wowacdm (right) vanilla analyses. The four shades within each group show the four redshift bins, in $\Mpch$ (red) and $\Mpc$ (blue) units; the dashed vertical lines mark zero deviation for each model, with the \wowacdm group shifted horizontally for visual clarity.}
    \label{fig:nuisance_zbins_whisker}
\end{figure}

The same preference for $\Mpc$ units emerges when looking at the nuisance parameters of the model. Figure~\ref{fig:nuisance_zbins_whisker} shows the weighted posterior mean of all the nuisance parameters, for each redshift bin of both the \wcdm and \wowacdm vanilla analyses, expressed as a deviation from the fiducial value in units of the corresponding posterior $1\sigma$ width. Across nearly all parameters and both models, the $\Mpc$ case is consistent with zero deviation, while the $\Mpch$ case shows persistent, often sizeable offsets. This is most evident for the linear bias $\bone$, whose systematic displacement by several $\sigma$ in $\Mpch$ units is particularly concerning given that it enters the model at leading order; this shift is also clearly correlated with the underestimation of $\As$ discussed above, the two parameters being strongly degenerate. The only exception to this otherwise clean separation is the next-to-leading-order counterterm $\cnlo$, for which a mild drift also starts to appear in $\Mpc$ units. Some residual \pve therefore persist even in $\Mpc$ units, and, as shown in Fig.~\ref{fig:comp_units_allcases_evo}, can be further reduced by resorting to the additional strategies discussed in Sect.~\ref{sec:results}.

\begin{figure*}
    \includegraphics[width=2\columnwidth]{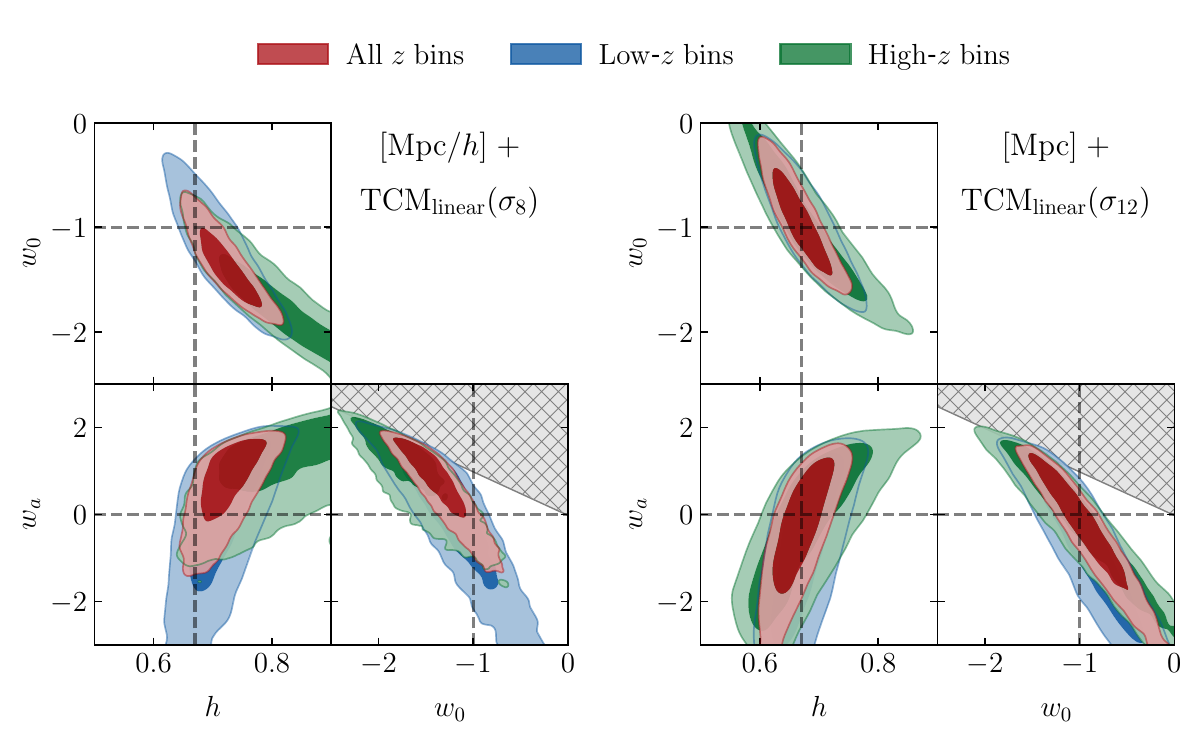}
    \caption{Comparison of the marginalised posterior distributions in the \wowacdm parameter space, obtained with the ${\rm TCM_{linear}}$ reparametrisation, calibrated on $\sigma_8$ in $\Mpch$ units (left) and on $\sigtwelve$ in $\Mpc$ units (right), using the DESI-like synthetic data described in App.~\ref{app:desi_like}. Dark red contours show the result of fitting all six redshift bins jointly; blue and green contours show the result of restricting the fit to the three lowest-redshift (BGS, LRG1, LRG2) and three highest-redshift (LRG3+ELG1, ELG2, QSO) tracers, respectively. The hatched region indicates $\wo+\wa>0$, excluded by the likelihood.}
    \label{fig:comp_DESIlike_zbins}
\end{figure*}

As anticipated in Sect.~\ref{sec:parametrisation}, Fig.~\ref{fig:comp_units_allcases_wCDM_w0waCDM} also allows us to test the impact of rescaling the stochastic contribution $\Pggnoise$ by the Alcock-Paczynski volume factor $A_{\rm AP}$ of Eq.~\eqref{eq:multipoles}, contrary to what we do throughout the rest of this work. The dashed ($\Mpch$) and dot-dashed ($\Mpc$) contours show the result of this customary rescaling, applied in addition to our adopted baseline treatment (filled contours). Rescaling the stochastic contribution by $A_{\rm AP}$ induces markedly larger prior-volume effects in $\Mpch$ units, where the resulting contours are shifted well beyond our adopted baseline, most noticeably along the $\wo$--$\wa$ degeneracy direction. This is much less relevant in $\Mpc$ units, where the two treatments yield well-aligned contours, consistent with the overall smaller impact of \pve in this unit convention. Although omitting the $A_{\rm AP}$ rescaling of the stochastic term is not by itself sufficient to remove \pve in $\Mpch$ units, we adopt this setup throughout the main body of this paper for the reasons discussed in Sect.~\ref{sec:parametrisation}. We note that the performance gap between the two unit conventions would be even larger had we instead adopted the more traditional approach of rescaling the stochastic contribution by $A_{\rm AP}$.

\section{Comparison of the performances of the TCM reparametrisation using DESI-like synthetic data}
\label{app:desi_like}

\begin{table}
\caption{Redshift bins, effective redshifts, effective sky areas, and number densities adopted to generate the DESI-like synthetic data set used in this appendix, representative of the BGS, LRG1, LRG2, LRG3+ELG1, ELG2, and QSO tracers.}
  \vspace{6pt}
  \centering
  \renewcommand{\arraystretch}{1.3}
  \begin{tabular}{|c|c|c|c|c|}
    \hline
    Tracer & $z$ range & $z_{\rm eff}$ & Area $[\deg^2]$ & $\bar n\,[h^3\,{\rm Mpc}^{-3}]$ \\
    \hline
    BGS & $[0.1,0.4]$ & $0.2953$ & $7180$ & $1.9\times10^{-4}$ \\
    LRG1 & $[0.4,0.6]$ & $0.5096$ & $5235$ & $2.1\times10^{-4}$ \\
    LRG2 & $[0.6,0.8]$ & $0.7059$ & $5235$ & $2.1\times10^{-4}$ \\
    LRG3+ELG1 & $[0.8,1.1]$ & $0.9184$ & $5235$ & $1.1\times10^{-4}$ \\
    ELG2 & $[1.1,1.6]$ & $1.3170$ & $4135$ & $1.0\times10^{-4}$ \\
    QSO & $[0.8,2.1]$ & $1.4901$ & $7077$ & $2.2\times10^{-5}$ \\
    \hline
  \end{tabular}
  \label{tab:desi_zbins}
\end{table}

To investigate why the $\Mpch$+${\rm TCM_{linear}}(\sigma_8)$ results in the main body of this paper are somewhat less optimal than those reported in \citep{TseCarMor2026}, likely due to differences in the redshift binning and survey configuration adopted, we generate a new set of synthetic data vectors reproducing the clustering measurements of the DESI DR1 analysis \citep{DESI_FS}. This setup comprises six redshift bins, representative of the BGS, LRG1, LRG2, LRG3+ELG1, ELG2, and QSO tracers, with the redshift ranges, effective redshifts, effective sky areas, and number densities listed in Table~\ref{tab:desi_zbins}. The associated covariance matrix is generated assuming once again the Gaussian approximation. As in the rest of this work, no systematic contribution is included in either case.

Figure~\ref{fig:comp_DESIlike_zbins} shows the resulting marginalised posterior distributions in the \wowacdm parameter space, comparing the $\Mpch$+${\rm TCM_{linear}}(\sigma_8)$ (left) and $\Mpc$+${\rm TCM_{linear}}(\sigtwelve)$ (right) reparametrisations. In $\Mpch$ units, we find a significant dependence of the recovered posterior on which redshift bins are included in the fit. Using only the low-redshift tracers (BGS, LRG1, LRG2), where the imprint of a varying dark-energy equation of state is stronger, the contours recover the fiducial \lcdm cosmology within $1\sigma$, although still exhibiting the premature breaking of the $\wo$--$\wa$ degeneracy already discussed in the main text, consistent with the findings of \citep{TseCarMor2026}. Using only the high-redshift tracers (LRG3+ELG1, ELG2, QSO), by contrast, we recover the same preference for low $\wo$ and high $\wa$ found in the main-body analysis, with the fiducial values only marginally consistent, between $2\sigma$ and $3\sigma$: at these redshifts, the likelihood is considerably shallower along the $(\wo,\wa)$ directions relative to the nuisance parameter space than in the low-redshift case, making the reparametrisation less effective at removing the associated \pve. Fitting all six redshift bins jointly moderately improves the agreement with the fiducial cosmology, but the result remains more biased than in the low-redshift-only case, partly on account of its greater constraining power. In $\Mpc$ units, on the other hand, the ${\rm TCM_{linear}}(\sigtwelve)$ reparametrisation performs consistently well regardless of which redshift bins are included, reinforcing our conclusion that adopting $\Mpc$ units is the more robust mitigation strategy.

\section{Robustness to the presence of Alcock-Paczynski distortions in the data}
\label{app:withAP}

Throughout this work, our synthetic data vectors were generated assuming perfect knowledge of the fiducial cosmology, such that no \apdist distortions are present in the data (i.e. $\qperp=\qpara=1$ by construction). This is an idealised setting: when analysing real data, the true cosmology is unknown, and the fiducial cosmology adopted to generate model predictions and perform unit conversions generally differs from the true one, so that \apdist distortions are imprinted in the data. To test the robustness of our conclusions against this more realistic setting, we generate two new sets of synthetic data vectors, both adopting a different cosmology from the fiducial one reported in Table~\ref{tab:cosmology}, while retaining the same reference cosmology used throughout the rest of this work to perform unit conversions and define priors; the resulting mismatch induces \apdist distortions in the synthetic data vectors. In both cases we adopt $h=0.8$, deliberately far from the fiducial value, to test whether it is correctly recovered, together with a correspondingly lower $\omc=0.11$---as opposed to the fiducial $\omc=0.121203$---for the same reason. We further adopt $\omb=0.02$ and $\ns=0.97$, on which we impose the same Gaussian priors used throughout the main analysis, centred on their respective fiducial values. The two data sets differ only in the assumed dark-energy equation of state: the first, hereafter AP$_1$, adopts $\wo=-0.8$ and $\wa=-1.0$, broadly consistent with recent DESI BAO and full-shape results \citep{AbdAguAhl2025, DESI_FS}; the second, hereafter AP$_2$, adopts $\wo=-1.2$ and $\wa=-0.5$, chosen to lie away from the $\wo$--$\wa$ degeneracy direction explored by AP$_1$.

\begin{figure*}
    \includegraphics[width=2\columnwidth]{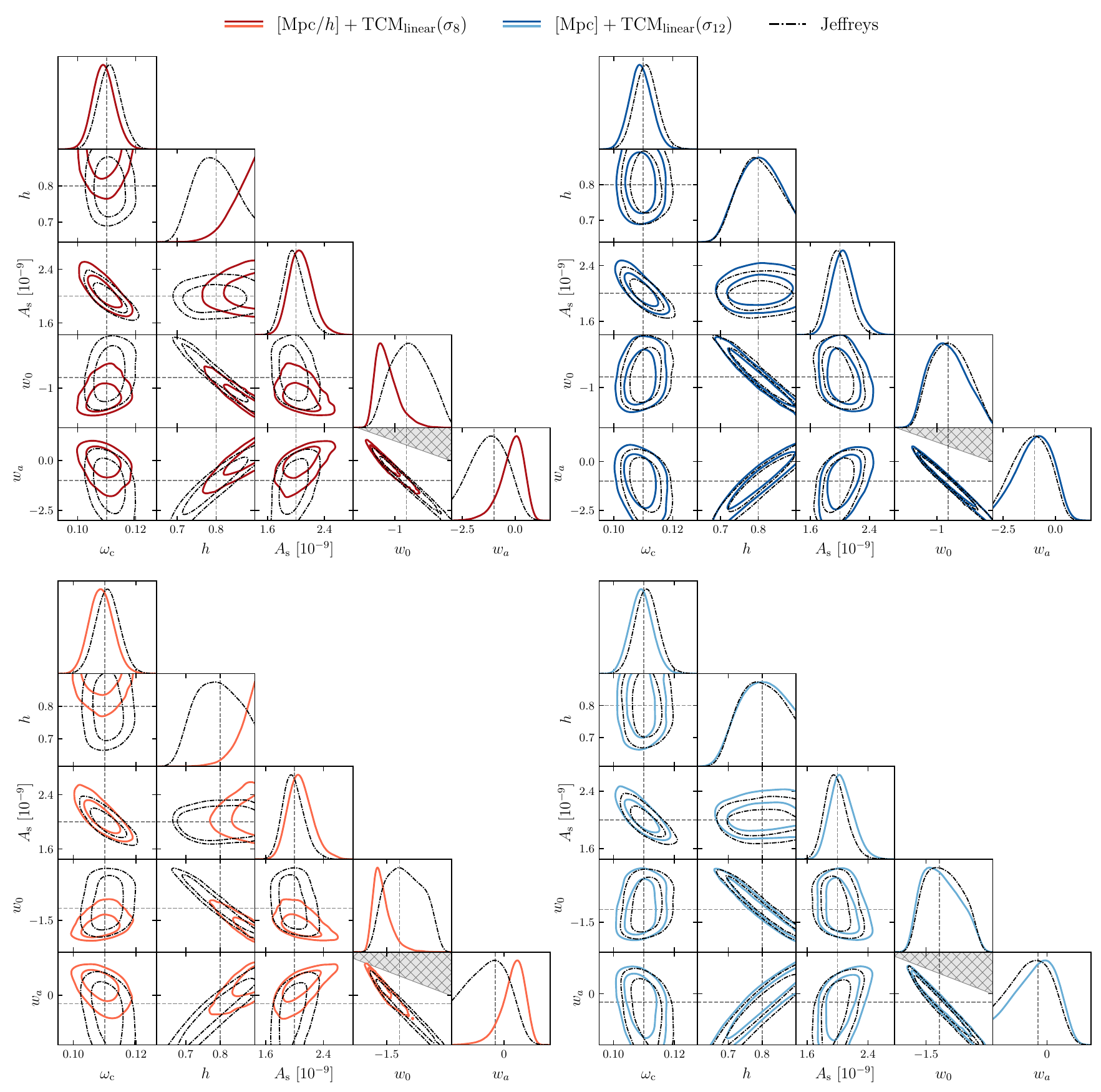}
    \caption{Comparison of the marginalised posterior distributions in the \wowacdm parameter space, using synthetic data vectors that include \apdist distortions, for the two cosmologies described above: AP$_1$ (top row) and AP$_2$ (bottom row). The left column shows results obtained in $\Mpch$ units, calibrated on $\sigma_8$; the right column shows the corresponding results in $\Mpc$ units, calibrated on $\sigtwelve$. In each panel, coloured contours show the ${\rm TCM_{linear}}$ reparametrisation (dark red/blue for AP$_1$, light red/blue for AP$_2$) and black dot-dashed contours show the results obtained with Jeffreys priors. The hatched region indicates $\wo+\wa>0$, excluded by the likelihood.}
    \label{fig:triangle_grid_withAP}
\end{figure*}

Figure~\ref{fig:triangle_grid_withAP} shows that, even in the presence of \apdist distortions directly imprinted in the data, we obtain results fully consistent with those found in the AP-distortion-free case discussed in Sect.~\ref{sec:reparam_schemes}, for both cosmologies considered: the $\Mpc$+${\rm TCM_{linear}}(\sigtwelve)$ combination remains consistently the best-performing configuration, closely tracking the Jeffreys reference case regardless of the assumed $(\wo,\wa)$, while the $\Mpch$+${\rm TCM_{linear}}(\sigma_8)$ combination is unable to fully correct for \pve in either case.

\clearpage
\bibliography{main}

\end{document}